\documentclass[aps,prd,twocolumn,nofootinbib,superscriptaddress,tightenlines]{revtex4}
\usepackage[dvipsnames]{xcolor}
\usepackage{amsmath}
\usepackage{dcolumn}
\usepackage{lipsum}
\usepackage{amssymb}
\usepackage{soul}
\usepackage{url}
\usepackage{epsfig}
\usepackage{graphicx}
\usepackage{amsmath}
\usepackage{bm}
\usepackage{setspace}
\usepackage{appendix}
\usepackage{lscape}
\usepackage{amsthm}
\usepackage{bbold}
\usepackage{dcolumn}
\usepackage{epsfig}
\usepackage{graphics}
\usepackage{graphicx}
\usepackage[utf8]{inputenc}

\usepackage{natbib}
\usepackage{graphicx}
\usepackage{dcolumn}
\usepackage{bm}
\usepackage{amsmath}
\usepackage{float}
\usepackage{multirow}
\usepackage{slashed}
\usepackage{xcolor}
\usepackage{physics}
\usepackage{multirow}
\usepackage{gensymb}
\usepackage{mathtools,braket}
\usepackage{subcaption}
\usepackage{lipsum}  
\usepackage{color}
\usepackage{soul}
\usepackage[colorlinks=true,
            linkcolor=red,
            citecolor=blue,
            urlcolor=blue]{hyperref}
\usepackage{bm}
\usepackage{xspace}
\usepackage{cancel}
\usepackage{float}
\usepackage{multirow}
\definecolor{darkgreen}{rgb}{0,0.5,0}
\definecolor{purple}{rgb}{0.5,0,0.5}
\definecolor{nblue}{rgb}{0.0,0.0,0.50}
\definecolor{scarlet}{rgb}{1.0,0.2,0}
\definecolor{darkmagenta}{rgb}{0.55, 0.0, 0.55}
\definecolor{darkolivegreen}{rgb}{0.33, 0.42, 0.18}
\definecolor{darkcandyapplered}{rgb}{0.64, 0.0, 0.0}
\usepackage[colorlinks=true]{hyperref}
\newcommand{\orcid}[1]{\href{https://orcid.org/#1}{\includegraphics[width=8pt]{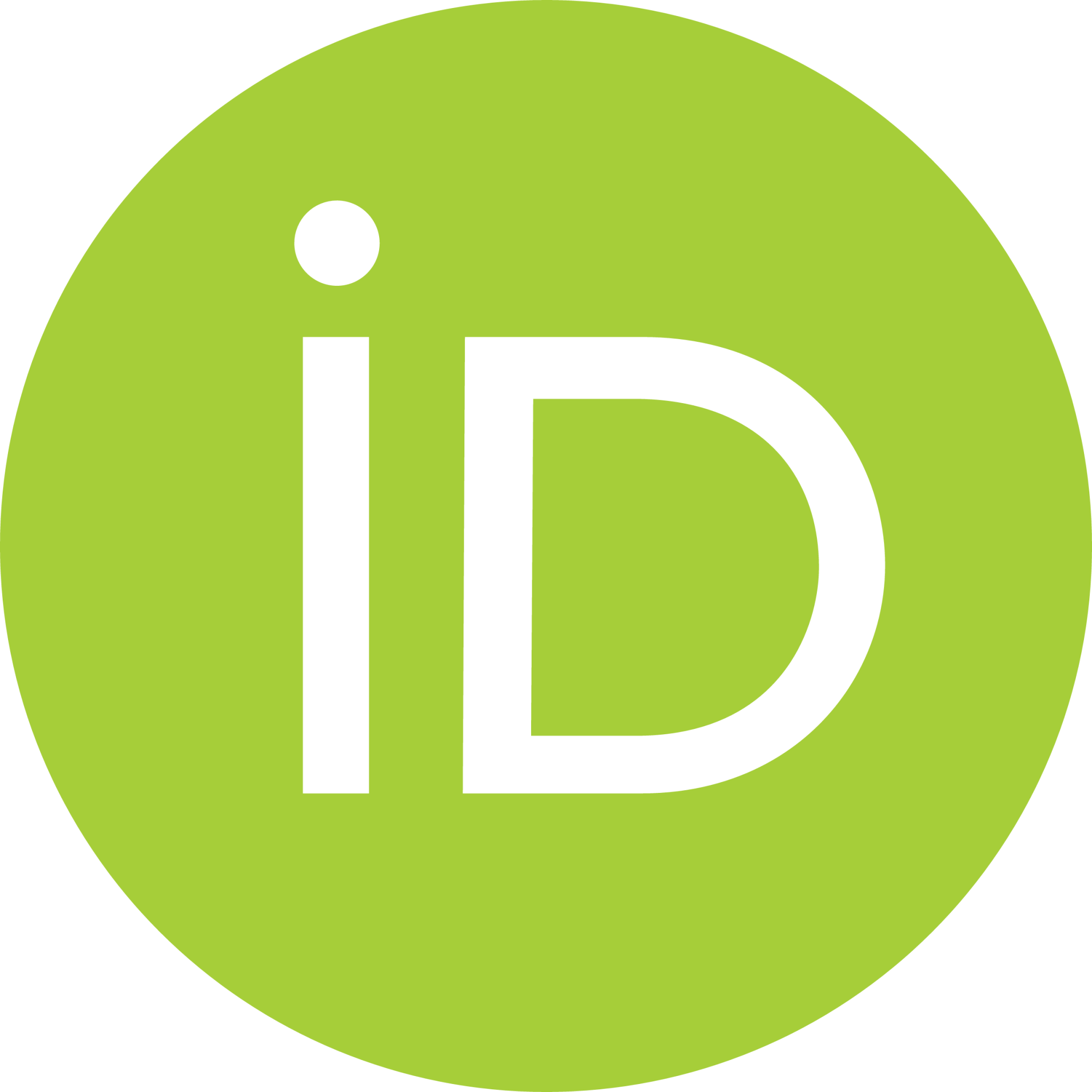}}}

\newcommand{\be}{\begin{equation}}

\newcommand{\bfk}{\textcolor{black}{\textbf{k}^2_\perp}}

\newcommand{\ee}{\end{equation}}
\newcommand{\bea}{\begin{eqnarray}}
\newcommand{\eea}{\end{eqnarray}}
\newcommand{\beas}{\begin{eqnarray*}}
\newcommand{\eeas}{\end{eqnarray*}}

\begin{document}
\title{An Analysis on the Parton Distribution Functions of Heavy Mesons}

\author{Satyajit Puhan\,\orcid{0009-0004-9766-5005}}
\email{puhansatyajit@gmail.com}
\affiliation{Computational High Energy Physics Lab, Department of Physics,
Dr.\ B.R.\ Ambedkar National Institute of Technology, Jalandhar, Punjab 144008, India}
\affiliation{Institute of Physics, Academia Sinica, Taipei 11529, Taiwan}

\begin{abstract}
In this work, we investigate the constituent parton distribution functions (PDFs) of the kaon and heavy pseudoscalar mesons within the light-cone quark model. Starting from the initial-scale quark and antiquark PDFs, obtained by evaluating the quark--quark correlation functions for individual mesons, we perform quantum chromodynamics (QCD) evolution to determine their partonic structure at higher energy scales. The QCD evolution is carried out using the next-to-leading-order (NLO) Dokshitzer--Gribov--Lipatov--Altarelli--Parisi (DGLAP) equations. We further compute the average longitudinal momentum fractions carried by the individual constituents at both the model and evolved scales. In addition, we predict the NLO structure functions of the kaon at energy scales relevant to the upcoming Electron--Ion Collider (EIC). For the COMPASS++/AMBER experiment, we also present detailed predictions for the NLO Drell--Yan cross sections induced by both $K^{+}$ and $K^{-}$ beams, using carbon, tungsten, and aluminum as nuclear targets. Finally, we demonstrate the dominance of the heavier constituents over the lighter constituents in heavy mesons in terms of the momentum fractions they carry.

\end{abstract}

\maketitle
\vspace{0.5em}

 \section{Introduction}
\label{intro}
It has long been a challenging task to describe the internal structure of hadrons in terms of their constituent quarks, antiquarks, gluons, and sea quarks. These constituents are primarily responsible for the internal dynamics of hadrons and their interactions with other hadrons. The strong interactions among the partons inside a hadron and the hadronic matrix elements of quark--gluon field operators can be understood within the framework of quantum chromodynamics (QCD) \cite{Brodsky:1997de}. QCD factorization theorems separate the long-distance non-perturbative contributions from the short-distance perturbative components in experimental cross sections, thereby enabling the study of hadron structure. However, the theoretical description of hadronic structure in the perturbative regime remains highly non-trivial because of the complex dynamics of valence quarks, sea quarks, and gluons. Consequently, the distributions of these constituents are investigated through quark--gluon correlation functions in the non-perturbative regime using low-energy effective models \cite{Mineo:2003vc,Nambu:1961tp,Klevansky:1992qe,Schlumpf:1994bc,Brodsky:2006uqa}. Owing to the non-perturbative nature of QCD, particularly color confinement and spontaneous chiral symmetry breaking, it remains difficult to describe hadron structure directly from first principles and to perform calculations based on the fundamental QCD Lagrangian \cite{Roberts:2000aa,Ji:2026lyj}. The distribution functions of the constituents encode information about various multidimensional degrees of freedom, including their momentum and spatial distributions. These include generalized transverse momentum-dependent distributions (GTMDs) \cite{Meissner:2009ww,Goeke:2005hb,Puhan:2025kzz}, generalized parton distributions (GPDs) \cite{Diehl:2003ny,Chavez:2021llq,Zhang:2021tnr,Broniowski:2022iip,Guidal:2004nd,Puhan:2026pkz}, transverse momentum-dependent distributions (TMDs) \cite{Diehl:2015uka,Angeles-Martinez:2015sea,Pasquini:2008ax,Puhan:2023ekt,Puhan:2023hio,Puhan:2025ujg}, parton distribution functions (PDFs) \cite{Collins:1981uw,Martin:1998sq,Gluck:1994uf}, form factors (FFs) \cite{Puhan:2025pfs,Arrington:2006zm}, and many others.

\par The one-dimensional PDFs are among the most fundamental non-perturbative inputs in QCD, encoding the longitudinal momentum structure of quarks and gluons inside hadrons. Determining the PDFs of different hadrons through hard-scattering processes is one of the central goals of hadron physics. As functions of the longitudinal momentum fraction $x$, PDFs describe the probability of finding a quark or gluon carrying a fraction $x$ of the parent hadron's longitudinal momentum \cite{Barry:2018ort}. Within the framework of QCD factorization, they provide a crucial link between the underlying partonic dynamics and experimentally measurable cross sections. PDFs can be extracted from the long-distance contributions to processes such as deep-inelastic scattering (DIS) \cite{Bloom:1969kc,Altarelli:1977zs}, leading-neutron electroproduction \cite{H1:2010hym,ZEUS:2002gig}, prompt photon production \cite{WA70:1987bai}, $J/\psi$ production \cite{Chang:2022pcb}, and Drell--Yan processes \cite{Drell:1970wh}. Although considerable theoretical and experimental efforts have been devoted to determining the PDFs of baryons, particularly nucleons \cite{Lorce:2025aqp,ATLAS:2021qnl}, comparatively little is known about the partonic structure of mesons. Among the mesons, the pion and the kaon play a central role as the lightest quark--antiquark bound states and the pseudo-Goldstone bosons associated with spontaneous chiral symmetry breaking \cite{Nambu:1961tp}. In our recent work \cite{P:2026crg}, we investigated the pion PDFs and presented predictions for upcoming EIC experiments. In the present work, we extend our PDF analysis to the kaon and other heavy pseudoscalar mesons within the light-cone quark model (LCQM).

\par Considerable theoretical model predictions \cite{Lan:2019rba,Gutsche:2014zua,Shi:2026hqq,P:2026crg}, experimental measurements \cite{E615:1989bda}, theoretical extractions \cite{Bourrely:2022mjf,Barry:2021osv,Chang:2020rdy,Novikov:2020snp,Gluck:1999xe}, and lattice QCD studies \cite{Miller:2025wgr,ExtendedTwistedMass:2024kjf} are available for pion PDFs compared with those of other mesons. In contrast, exploring the PDFs of the kaon and other mesons experimentally remains challenging. Nevertheless, their partonic structure can be investigated through theoretical models and lattice QCD simulations. For the kaon, the constituent PDFs have been studied within light-front quantization \cite{Lan:2019rba}, the light-front quark model (LFQM) \cite{Choi:2025bxk}, the light-front holographic model (LFHM) \cite{deTeramond:2008ht}, the Dyson--Schwinger equations (DSE) framework \cite{Ding:2019lwe}, and the chiral quark model \cite{Nam:2012vm}. Kaon PDFs have also been investigated using lattice QCD \cite{Barry:2025wjx,Miller:2025wgr} and through global theoretical extractions \cite{Chang:2024rbs,Bourrely:2023yzi}. Experimentally, only a single dataset for the kaon-induced Drell--Yan process is available, obtained nearly four decades ago by the NA3 collaboration \cite{oter}. The measurements revealed that the $\bar{u}$ quark distribution in the $K^{-}$ meson is softer than that in the $\pi^{-}$ meson, reflecting the suppression of the $\bar{u}$ quark due to the presence of the heavier $s$ quark and consequently the breaking of SU(3) flavor symmetry. Apart from fixed-target Drell--Yan measurements, kaon beams have also been used to study $J/\psi$ production \cite{Corden:1980rb}. The ongoing COMPASS++/AMBER experiment is expected to provide high-precision data for the kaon-induced Drell--Yan process using carbon targets through the reaction $K^-+C\rightarrow\mu^+\mu^-+X$ \cite{Adams:2018pwt}. Furthermore, the future Electron--Ion Collider (EIC) will probe kaon PDFs through tagged deep-inelastic scattering (DIS) and the Sullivan process, $e+p\rightarrow e'+X+\Lambda$ \cite{Lu:2025bnm,Aguilar:2019teb}. Heavy-meson PDFs are even more difficult to measure experimentally because of the extremely short lifetimes of heavy mesons. Nevertheless, heavy mesons provide an ideal laboratory for investigating the interplay between perturbative and non-perturbative QCD due to the presence of heavy quarks. Their constituent quarks move more slowly inside the bound state than those in light mesons, making them particularly interesting systems for studying heavy-flavor dynamics. PDFs of heavy mesons have been investigated within several theoretical frameworks \cite{Wu:2025rto,Albino:2022gzs,Lan:2019img,Tang:2019gvn,Shi:2024laj,Puhan:2024ckp}. A detailed understanding of heavy-meson PDFs is essential for describing hard-scattering processes, heavy-flavor production in high-energy collisions, and precision predictions for electroweak boson and Higgs production, as well as collider phenomenology \cite{Lan:2019img}.

\par In this work, we calculate the valence quark and antiquark PDFs of the kaon and several heavy pseudoscalar mesons by evaluating the quark--quark correlation functions within the LCQM. Since higher Fock states contribute only marginally for heavy quarks \cite{Shi:2022erw}, we restrict our analysis to the lowest quark--antiquark Fock state. The LCQM provides a relativistic and gauge-invariant framework for describing hadron structure and properties, including mass spectra, decay constants, and radiative decays. Owing to the boost-invariant nature of its wave functions, the LCQM offers a consistent description of hadrons in different reference frames. The model primarily describes the valence-quark structure of hadrons and has been shown to provide reliable predictions even after QCD evolution to perturbative scales. For pseudoscalar mesons, only the leading-twist collinear distribution $f(x)$ exists, unlike the nucleon case, where three leading-twist PDFs are present. The distribution $f(x)$ describes the probability of finding an unpolarized quark inside an unpolarized hadron. By evaluating the quark--quark correlation function, we derive the PDFs both in terms of the light-front wave functions (LFWFs) and in their explicit analytical forms using the complete meson wave function, including both spin and momentum components. At the model scale, the PDFs contain only valence-quark contributions, while the gluon and sea-quark distributions vanish. We consider charm- and bottom-quark PDFs for the $D$- and $B$-meson families as well as for heavy quarkonia. The initial-scale PDFs are evolved to higher energy scales using the next-to-leading-order (NLO) Dokshitzer--Gribov--Lipatov--Altarelli--Parisi (DGLAP) evolution equations \cite{Karlberg:2025hxk}. We determine the valence, gluon, and sea-quark PDFs together with their Mellin moments at different energy scales. In addition, we predict the kaon structure functions and the NLO fixed-target Drell--Yan cross sections. For the Drell--Yan calculations, we employ carbon, aluminum, and tungsten nuclear PDFs available through the LHAPDF library \cite{Buckley:2014ana} and consider both $K^{+}$- and $K^{-}$-induced reactions. We further determine the average momentum fractions carried by the valence quarks, gluons, and sea quarks at different energy scales and evaluate higher-order Mellin moments. Our predictions for the average momentum fractions in the kaon are compared with available theoretical and lattice QCD results. We also investigate the dominance of heavy constituents over light constituents in heavy mesons, the quark--antiquark flavor asymmetry in different mesons, and the behavior of identical quark flavors across different mesonic systems. These studies provide further insight into the gluon and sea-quark structure of heavy mesons.

\par The paper is organized as follows. In Sec.~\ref{satya}, we introduce the LCQM and discuss its spin and momentum wave functions. In Sec.~\ref{pdf}, we present the valence quark and antiquark PDFs at the model scale, derive the explicit expressions for the LFWFs and their Mellin moments, and predict the NLO kaon structure functions and unpolarized Drell--Yan cross sections. We also discuss the flavor asymmetry at both the model and evolved scales for different mesons. Finally, our conclusions are presented in Sec.~\ref{conclus}.
 \section{Methodology}
 \label{satya}
 \subsection{Light-cone quark model}

The hadron wave function based on the light-front (LF) quantization of QCD using the multi-particle Fock-state expansion is expressed as \cite{Hecht:2000xa,Puhan:2023ekt,Lepage:1980fj,Ji:2003yj,Brodsky:2000xy,Brodsky:2000dr,Pasquini:2023aaf,Puhan:2025kzz,Brodsky:1997de}
\begin{equation}
\begin{aligned}
|\Psi_h (P^+, \mathbf{P}_\perp, S_z) \rangle_\Lambda
=& \sum_{n,\lambda_i}
\int \prod_{i=1}^n
\frac{\mathrm{d} x_i \, \mathrm{d}^2 \mathbf{k}_{\perp i}}
{\sqrt{x_i}\,16\pi^3}
\\
&\times 16\pi^3
\delta\!\left(1-\sum_{i=1}^n x_i\right)
\delta^{(2)}\!\left(\sum_{i=1}^n \mathbf{k}_{\perp i}\right)
\\
&\times
\psi^{\Lambda}_{n/h}(x_i,\mathbf{k}_{\perp i},\lambda_i)
\\
&\times
| n ; x_i P^+,\, x_i \mathbf{P}_\perp+\mathbf{k}_{\perp i},
\lambda_i \rangle .
\end{aligned}
\label{fockstate}
\end{equation}

Here, $|\Psi_h(P^+,\mathbf{P}_\perp,S_z)\rangle_\Lambda$ denotes the hadron eigenstate with the four-vector LF momentum $P=(P^+,P_\perp,P^-)$ and spin $\Lambda$. The quantity $S_z$ represents the spin projection of the hadron. For spin-0 pseudoscalar mesons, $S_z=\Lambda=0$. In the case of $\Lambda=1$ (vector mesons), $S_z$ takes the values $\pm1$ and $0$. The variables $x_i$ and $\mathbf{k}_{\perp i}$ denote the longitudinal momentum fraction and transverse momentum, respectively, of the $i^{\rm th}$ constituent of the hadron. The quantity $\lambda_i$ represents the helicity of the $i^{\rm th}$ constituent. The relation between $S_z$ and $\lambda_i$ is given by
$S_z=\sum_i \lambda_i+l_z$,
where $l_z$ is the orbital angular momentum of the hadron. In LF dynamics, the momentum and energy of the hadron are distributed among its constituents and satisfy the corresponding conservation laws.

The four-vector momentum $k_i$ of the $i^{\rm th}$ constituent can be expressed as
\begin{eqnarray}
k_i=(k^+_i,\textbf{k}_{\perp i},k_i^-)=(x_i P^+,\, x_i P_\perp +\textbf{k}_{\perp i},\,\frac{\textbf{k}_{\perp i}^2+m_i^2}{k^+_i} ).
\end{eqnarray}

Here, $m_i$ is the mass of the $i^{\rm th}$ constituent. The total longitudinal momentum carried by the constituents satisfies the momentum conservation relation,
$P^+=\sum_i^n k_i^+$,
which implies that
$\sum_i^n x_i=1$.
Similarly, the constituents satisfy energy conservation,
$P^-=\sum_i^n k_i^-$.
The transverse momenta also satisfy the conservation relation
$\sum_i^n\mathbf{k}_{\perp i}=0$.
The function $\psi^{\Lambda}_{n/h}(x_i,\mathbf{k}_{\perp i},\lambda_i)$ in Eq.~(\ref{fockstate}) represents the LF wave function, which is independent of the parent hadron momentum. The hadron eigenstate in Eq.~(\ref{fockstate}) obeys the normalization condition
\begin{align}
{}_{\Lambda'}\!\langle \Psi_h(P^{+\prime},\mathbf{P}'_\perp,S_z)
\mid
\Psi_h(P^+,\mathbf{P}_\perp,S_z) \rangle_\Lambda
\nonumber\\
=\,2(2\pi)^3 P^+\,\delta_{\Lambda'\Lambda}\,
\delta(P^+-P^{+\prime})
\delta^{(2)}(\mathbf{P}_\perp-\mathbf{P}'_\perp).
\end{align}

The above formalism is applicable to any composite system, irrespective of the reference frame. In the present work, however, we restrict our discussion to the meson state, which is the primary focus of this study.
\par The Fock state of a meson can be expressed in terms of quarks, gluons, and sea-quark degrees of freedom as \cite{Pasquini:2023aaf,Puhan:2023ekt,Luan:2024dvc}
\begin{equation}
\begin{aligned}
|\Psi_m\rangle
=& \sum |q\bar{q}\rangle\, \psi_{q\bar{q}}
+ \sum |q\bar{q}g\rangle\, \psi_{q\bar{q}g}
+ \sum |q\bar{q}gg\rangle\, \psi_{q\bar{q}gg}
\\
&+ \sum |q\bar{q}(q\bar{q})_{\rm sea}\rangle\,
\psi_{q\bar{q}(q\bar{q})_{\rm sea}}
+ \cdots
\end{aligned}
\end{equation}

In the present work, the Fock-state expansion of the meson wave function in the LCQM is restricted to the lowest Fock state. The spin-orbit component of the lowest Fock state is regarded as a free state, with its quantum numbers, including angular momentum, parity, and charge conjugation, incorporated into the light-front wave function through the Melosh--Wigner transformation \cite{Qian:2008px}. Therefore, the meson is treated as a constituent quark--antiquark system. Consequently, Eq.~(\ref{fockstate}) can be reduced to the meson eigenstate by considering only the two-particle Fock state ($n=2$) as \cite{Tanisha:2025qda,Puhan:2024jaw,Puhan:2023ekt,Brodsky:2000dr}

\begin{equation}
\begin{aligned}
|\Psi_h (P^+,\mathbf{P}_\perp,S_z)\rangle_\Lambda
=& \sum_{\lambda_1,\lambda_2}
\int \frac{\mathrm{d}x\,\mathrm{d}^2\mathbf{k}_\perp}
{\sqrt{x(1-x)}\,16\pi^3}
\\
&\times \psi^{\Lambda}_{q\bar q}
(x,\mathbf{k}_\perp,\lambda_1,\lambda_2)
\\
&\times \Big|
xP^+,\mathbf{k}_\perp,\lambda_1;
(1-x)P^+,-\mathbf{k}_\perp,\lambda_2
\Big\rangle .
\end{aligned}
\label{meson}
\end{equation}

Here, $\lambda_1$ and $\lambda_2$ denote the helicities of the quark and antiquark, respectively. Throughout this work, we use $q$ and $\bar q$ to represent the quark and antiquark, respectively. The four-momenta of the quark ($k_q$) and antiquark ($k_{\bar q}$) are given by

\begin{eqnarray}
	k_q&\equiv&\bigg(x P^+, \frac{\textbf{k}_\perp^2+m_q^2}{x P^+},\textbf{k}_\perp \bigg),\\
	k_{\bar q}&\equiv&\bigg((1-x) P^+, \frac{\textbf{k}_\perp^2+m_{\bar q}^2}{(1-x) P^+},-\textbf{k}_\perp \bigg).
\end{eqnarray}

The function $\psi^{\Lambda}_{q \bar q}(x,\mathbf{k}_{\perp},\lambda_1,\lambda_2)$ in Eq.~(\ref{meson}) represents the momentum-space light-front wave function (LFWF), which can be written as the product of the radial and spin wave functions \cite{Acharyya:2024enp}

\begin{eqnarray}
	\psi^{\Lambda}_{q \bar q}(x,\mathbf{k}_{\perp},\lambda_1,\lambda_2)=\phi(x,\bfk)\mathcal{S}_\Lambda(x,\mathbf{k}_\perp,\lambda_1,\lambda_2), \label{mesonwave}
\end{eqnarray}

where $\phi(x,\bfk)$ and $\mathcal{S}_\Lambda(x,\mathbf{k}_\perp,\lambda_1,\lambda_2)$ denote the radial and spin wave functions of the meson, respectively. For the ground-state meson momentum-space wave function, we adopt the BHL prescription \cite{Brodsky:1980vj,Brodsky:1980ny}. Within this prescription, the radial wave function of the meson is expressed as

\begin{align}
\phi(x,\mathbf{k}^2_\perp) =
A_h \exp \Bigg[
&-\frac{\frac{\mathbf{k}_\perp^2+m_q^2}{x}
+\frac{\mathbf{k}_\perp^2+m_{\bar q}^2}{1-x}}
{8\beta_h^2}+\frac{m_q^2+m^2_{\bar q}}{4 \beta_h^2}
\nonumber \\
&-\frac{(m_q^2-m_{\bar q}^2)^2}
{8\beta_h^2\left(
\frac{\mathbf{k}_\perp^2+m_q^2}{x}
+\frac{\mathbf{k}_\perp^2+m_{\bar q}^2}{1-x}
\right)}
\Bigg],
\label{bhl-k}
\end{align}

for asymmetric quark--antiquark masses \cite{Xiao:2002iv,Puhan:2025pfs}. Here, $m_q$ and $m_{\bar q}$ denote the quark and antiquark masses, respectively. For symmetric quark--antiquark masses ($m_q=m_{\bar q}$), the radial wave function reduces to \cite{Qian:2008px,Puhan:2025ibn}

\begin{eqnarray}
	\phi(x,\bfk)=A_h \ {\rm exp}\bigg[-\frac{1}{8 \beta_h^2} \frac{{\bf k}^2_\perp+m_{q(\bar q)}^2} {x(1-x)}+ \frac{m_{q(\bar q)}^2}{2 \beta_h^2} \bigg].
	\label{bhl-pi}
\end{eqnarray}

Here, $A_h$ and $\beta_h$ are the normalization constant and harmonic-scale parameter of the corresponding meson, respectively. The normalization constant $A_h$ is determined by normalizing the radial wave function $\phi(x,\bfk)$ according to

\begin{eqnarray}
	\int \frac{dx\, d^2\textbf{k}_\perp}{2 (2\pi)^3}|\phi(x,\bfk)|^2=1.
\end{eqnarray}

The values of the harmonic-scale parameter $\beta_h$, the quark mass $m_q$, and the antiquark mass $m_{\bar q}$ used for different mesons are listed in Table~\ref{tab1}. These parameters are adopted from our previous work \cite{Puhan:2024jaw,Puhan:2023ekt}, where they were determined by reproducing the meson masses using the variational principle \cite{Arifi:2022pal}.
\begin{table*}[t]
\centering
\caption{The constituent light quark masses $m_q$ ($q = u, d, s$) and heavy quark masses $m_Q$ ($Q = c, b$), along with the harmonic oscillator scale parameters $\beta_h$, are given in units of GeV. These parameters are adopted from Ref.~\cite{Arifi:2022pal,Puhan:2023ekt}.}
\label{tab1}
\begin{tabular}{|c|c|c|c|c|c|c|c|c|c|c|c|}
\hline
$m_{u(d)}$ & $m_s$ & $m_c$ & $m_b$ & $\beta_{qs}$ & $\beta_{qc}$ & $\beta_{cc}$ & $\beta_{bb}$ & $\beta_{sc}$ & $\beta_{qb}$ & $\beta_{sb}$ & $\beta_{bc}$ \\
\hline
0.22 & 0.45 & 1.68 & 5.10  & 0.405 & 0.500 & 0.699 & 1.376 & 0.537 & 0.585 & 0.636 & 0.906 \\
\hline
\end{tabular}
\end{table*}
Irrespective of the meson spin, the radial wave function remains unchanged. However, the spin wave function $\mathcal{S}_\Lambda(x,\mathbf{k}_\perp,\lambda_1,\lambda_2)$ in Eq.~(\ref{mesonwave}) differs for mesons with different spin quantum numbers. In the present work, we restrict our analysis to spin-0 pseudoscalar mesons ($\Lambda=S_z=0$). The front-form spin wave function can be derived either from the instant-form spin wave function through the Melosh--Wigner rotation \cite{Qian:2008px,Xiao:2002iv} or by solving the quark--meson vertex using the appropriate Dirac spinors. Both approaches lead to the same spin wave function for spin-0 pseudoscalar mesons. In this work, we employ the spin wave function obtained from the quark--meson vertex, following our previous studies \cite{Choi:1996mq,Qian:2008px,Dwibedi:2025vhr,Puhan:2025ujg,Puhan:2025kzz}. The spin wave function for spin-0 pseudoscalar mesons ($\Lambda=S_z=0$) is given by
\begin{align}
\mathcal{S}(x,\mathbf{k}_\perp,\lambda_1,\lambda_{2})
=&\,
\bar u(k_q,\lambda_1)
\notag\\
&\times
\frac{\mathcal{A}_{q \bar q}\gamma_5}
{\sqrt{2(M_{q \bar q}^2-(m_q^2-m_{\bar q}^2))}}
\, v(k_{\bar q},\lambda_2),
\end{align}
where $\mathcal{A}_{q \bar q}=M_{q \bar q}+m_q+m_{\bar q}$. Here, $u$ and $v$ denote the light-front Dirac spinors \cite{Harindranath:1996hq}, and
\[
M_{q \bar q}=\sqrt{\frac{\mathbf{k}_\perp^2+m_q^2}{x}+\frac{\mathbf{k}_\perp^2+m_{\bar q}^2}{1-x}}
\]
is the invariant mass of the quark--antiquark system. The spin wave functions for pseudoscalar mesons ($S_z=0$) corresponding to different helicity combinations of the quark and antiquark are given by \cite{Qian:2008px}
\begin{equation}
\begin{aligned}
\mathcal{S}(x,\mathbf{k}_\perp, \uparrow,\uparrow)
&= \frac{1}{\sqrt{2}}\omega^{-1}(-\mathbf{k}^{L})
\mathcal{A}_{q \bar q}, \\
\mathcal{S}(x,\mathbf{k}_\perp, \uparrow,\downarrow)
&= \frac{1}{\sqrt{2}}\omega^{-1}\big((1-x)m_q+x m_{\bar q}\big)
\mathcal{A}_{q \bar q}, \\
\mathcal{S}(x,\mathbf{k}_\perp, \downarrow,\uparrow)
&= \frac{1}{\sqrt{2}}\omega^{-1}\big(-(1-x)m_q-x m_{\bar q}\big)
\mathcal{A}_{q \bar q}, \\
\mathcal{S}(x,\mathbf{k}_\perp, \downarrow,\downarrow)
&= \frac{1}{\sqrt{2}}\omega^{-1}(-\mathbf{k}^{R})
\mathcal{A}_{q \bar q}.
\label{spin1}
\end{aligned}
\end{equation}
Here,
\[
\omega=\mathcal{A}_{q \bar q}\sqrt{x(1-x)\left[M_{q \bar q}^2-(m_q-m_{\bar q})^2\right]}.
\]
The spin wave function satisfies the normalization condition
\begin{equation}
\sum_{\lambda_1,\lambda_{2}}
\left|
\mathcal{S}(x,\mathbf{k}_\perp,\lambda_1,\lambda_2)
\right|^{2}
=1.
\end{equation}

Finally, the two-particle Fock state for spin-0 pseudoscalar mesons in Eq.~(\ref{meson}) can be written in terms of all possible helicity combinations of the constituent quark and antiquark together with the momentum-space wave function as
\begin{widetext}
\begin{equation}
\begin{aligned}
|\Psi_h (P^+,\mathbf{P}_\perp,S_z=0)\rangle
=&
\int
\frac{dx\, d^2\mathbf{k}_\perp}
{2(2\pi)^3 \sqrt{x(1-x)}}
\,\phi(x,\mathbf{k}_\perp^{2})
\\
&\times
\Big[
\mathcal{S}(x,\mathbf{k}_\perp,\uparrow,\uparrow)
|xP^+,\mathbf{k}_\perp,\uparrow;\,(1-x)P^+,-\mathbf{k}_\perp,\uparrow\rangle+
\mathcal{S}(x,\mathbf{k}_\perp,\downarrow,\downarrow)
|xP^+,\mathbf{k}_\perp,\downarrow;\,(1-x)P^+,-\mathbf{k}_\perp,\downarrow\rangle
\\
&\quad+
\mathcal{S}(x,\mathbf{k}_\perp,\downarrow,\uparrow)
|xP^+,\mathbf{k}_\perp,\downarrow;\,(1-x)P^+,-\mathbf{k}_\perp,\uparrow\rangle+
\mathcal{S}(x,\mathbf{k}_\perp,\uparrow,\downarrow)
|xP^+,\mathbf{k}_\perp,\uparrow;\,(1-x)P^+,-\mathbf{k}_\perp,\downarrow\rangle
\Big].
\label{eqeq}
\end{aligned}
\end{equation}
\end{widetext}

\section{Parton Distribution Functions}
\label{pdf}
\begin{figure}[t]
\centering
\includegraphics[width=\linewidth]{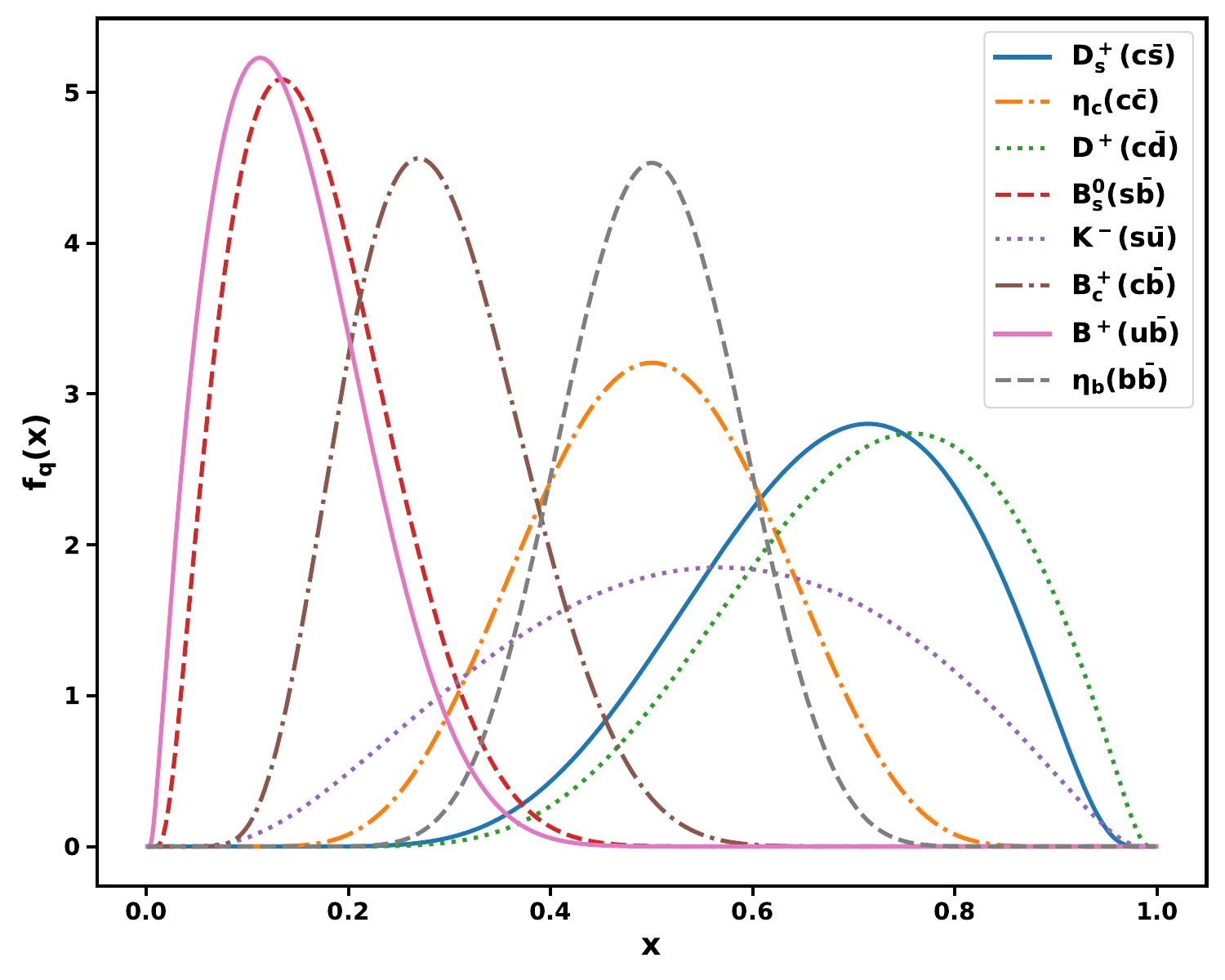}
\caption{(Color online) The quark PDFs, $f_q(x)$, of different mesons as functions of the longitudinal momentum fraction $x$ at their respective initial scales.}
\label{allpdf}
\end{figure}

\begin{table*}[t]
\centering
\caption{Calculated Mellin moments of the parton distribution functions at the initial scales, $\langle x^{n} \rangle$, for $n=0,1,2,3,$ and $4$. For quarkonium states, the quark and antiquark moments are identical, whereas they are listed separately for heavy--light and light--heavy mesons.}
\begin{ruledtabular}
\begin{tabular}{lcccccc}
\label{tab2}
Meson & Parton & $\langle x^{0} \rangle$ & $\langle x^{1} \rangle$
& $\langle x^{2} \rangle$ & $\langle x^{3} \rangle$ & $\langle x^{4} \rangle$ \\
\hline

$K^-(s \bar u)$ & $s$ & 1.000 & 0.551 & 0.338 & 0.224 & 0.157 \\
& $\bar u$ & 1.000 & 0.449 & 0.236 & 0.138 & 0.0861 \\

$D_s (c \bar s)$ & $c$ & 1.000 & 0.677 & 0.476 & 0.345 & 0.256 \\
& $\bar s$ & 1.000 & 0.323 & 0.121 & 0.051 & 0.023 \\
$D^+ (c \bar d)$ & $c$ & 1.000 & 0.714 & 0.528 & 0.402 & 0.313 \\
& $\bar d$ & 1.000 & 0.286 & 0.099 & 0.039 & 0.017 \\

$B_s^0(s \bar b)$ & $s$ & 1.000 & 0.166 & 0.033 & 0.008 & 0.002 \\
& $\bar b$ & 1.000 & 0.834 & 0.700 & 0.593 & 0.506 \\

$B^+(u \bar b)$ & $u$ & 1.000 & 0.145 & 0.027 & 0.006 & 0.001 \\
& $\bar b$ & 1.000 & 0.855 & 0.736 & 0.638 & 0.557 \\

$B_c^+(c \bar b)$ & $c$ & 1.000 & 0.290 & 0.091 & 0.031 & 0.011 \\
& $\bar b$ & 1.000 & 0.710 & 0.511 & 0.372 & 0.275 \\

$\eta_b(b \bar b)$ & $b/\bar b$ & 1.000 & 0.500 & 0.257 & 0.136 & 0.073 \\

$\eta_c(c \bar c)$ & $c/\bar c$ & 1.000 & 0.500 & 0.263 & 0.145 & 0.083 \\

\end{tabular}
\end{ruledtabular}
\end{table*}

\begin{figure*}[t]
\centering

\begin{subfigure}[t]{0.32\textwidth}
\centering
\vspace{0pt}
\includegraphics[width=\linewidth]{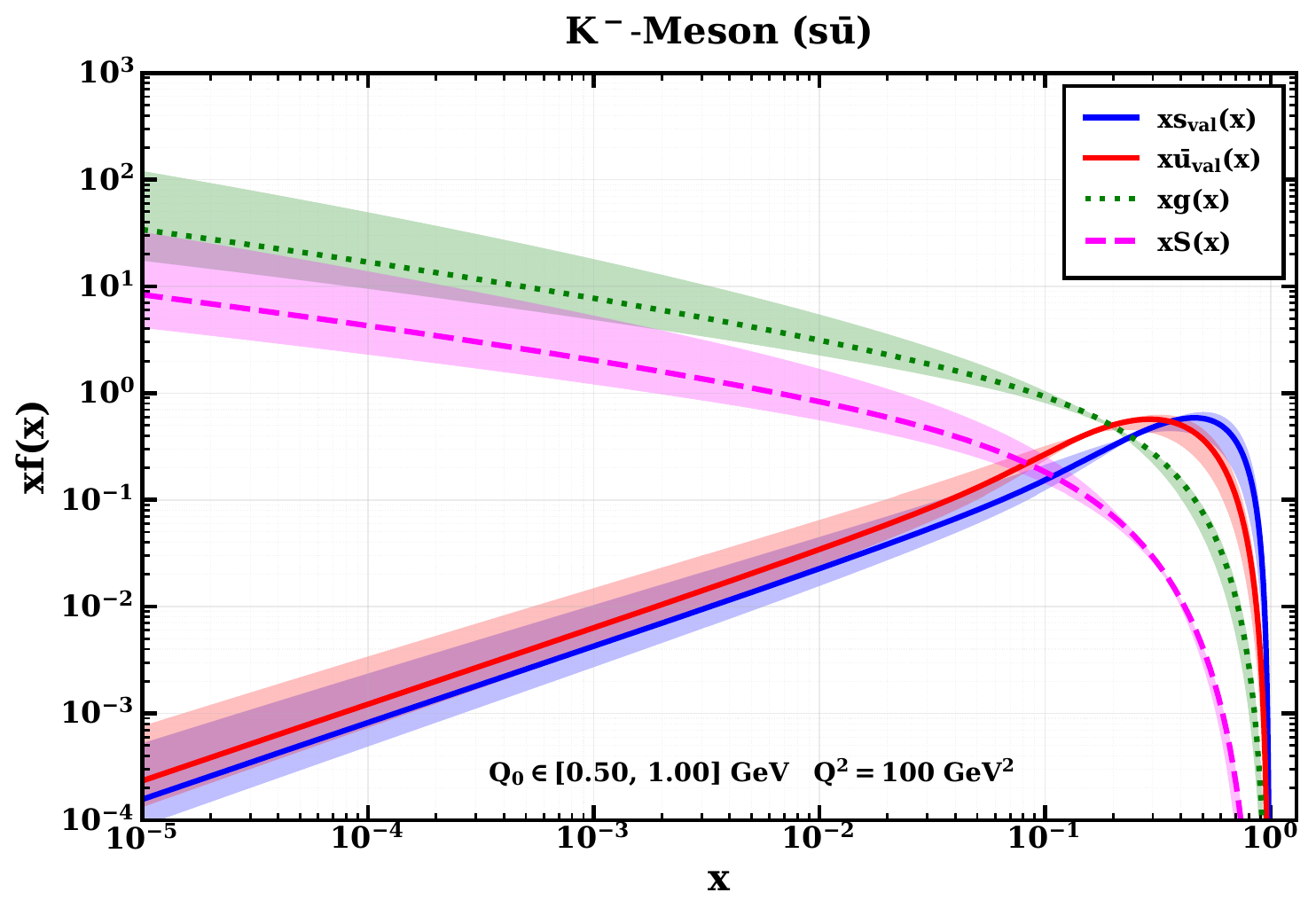}
\caption{The valence-quark distribution $xs_{\rm val}(x)$, valence antiquark distribution $x\bar{u}_{\rm val}(x)$, gluon distribution $xg(x)$, and sea-quark distribution $xS(x)$ as functions of the longitudinal momentum fraction $x$.}
\label{fig:kaon_pdf}
\end{subfigure}
\hfill
\begin{subfigure}[t]{0.32\textwidth}
\centering
\vspace{0pt}
\includegraphics[width=\linewidth]{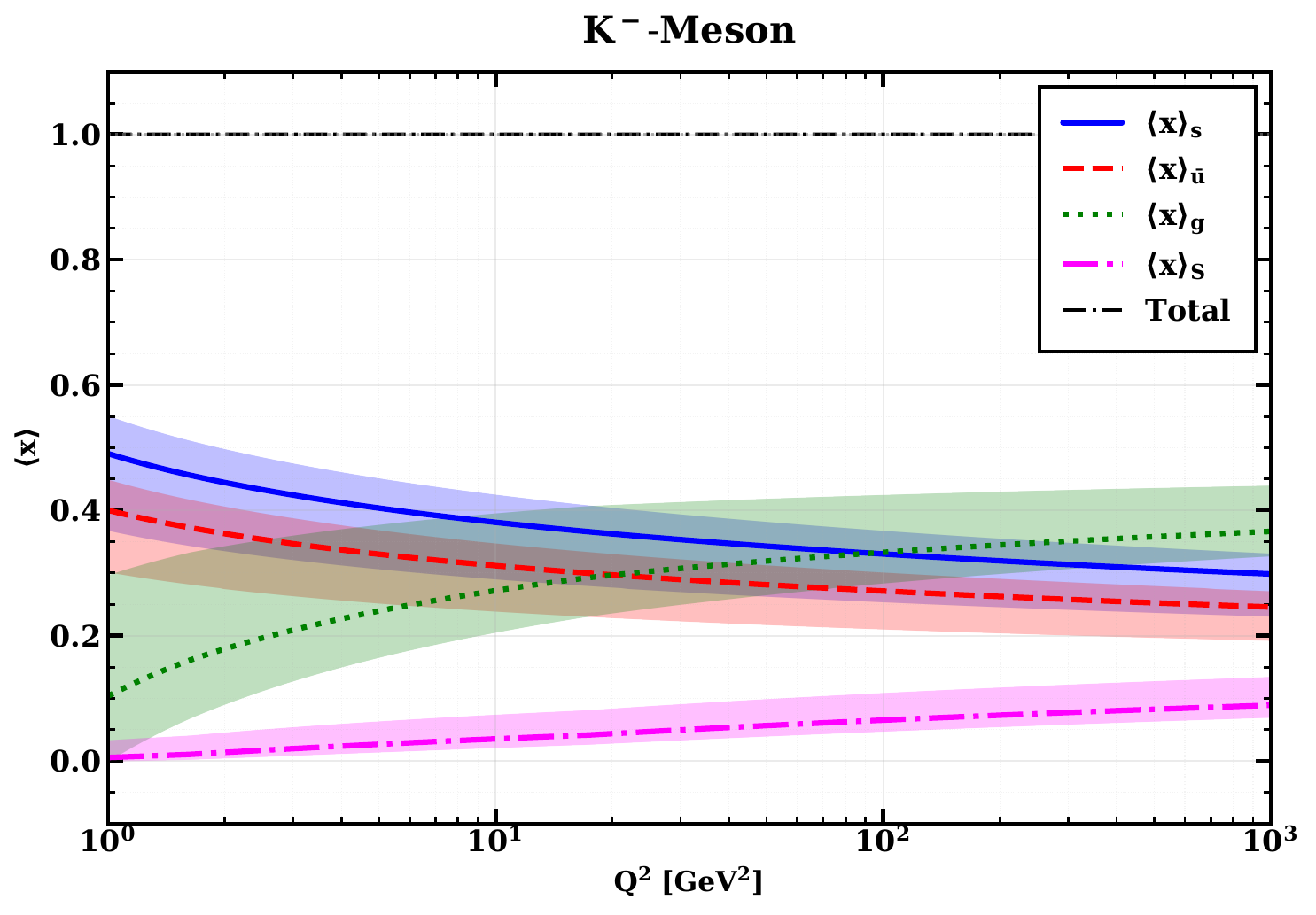}
\caption{Average momentum fractions carried by the valence quark, antiquark, gluon, and sea quarks, $\langle x\rangle_s$, $\langle x\rangle_{\bar u}$, $\langle x\rangle_g$, and $\langle x\rangle_S$, respectively, as functions of $Q^2$.}
\label{fig:kaon_momentum}
\end{subfigure}
\hfill
\begin{subfigure}[t]{0.32\textwidth}
\centering
\vspace{0pt}
\includegraphics[width=\linewidth]{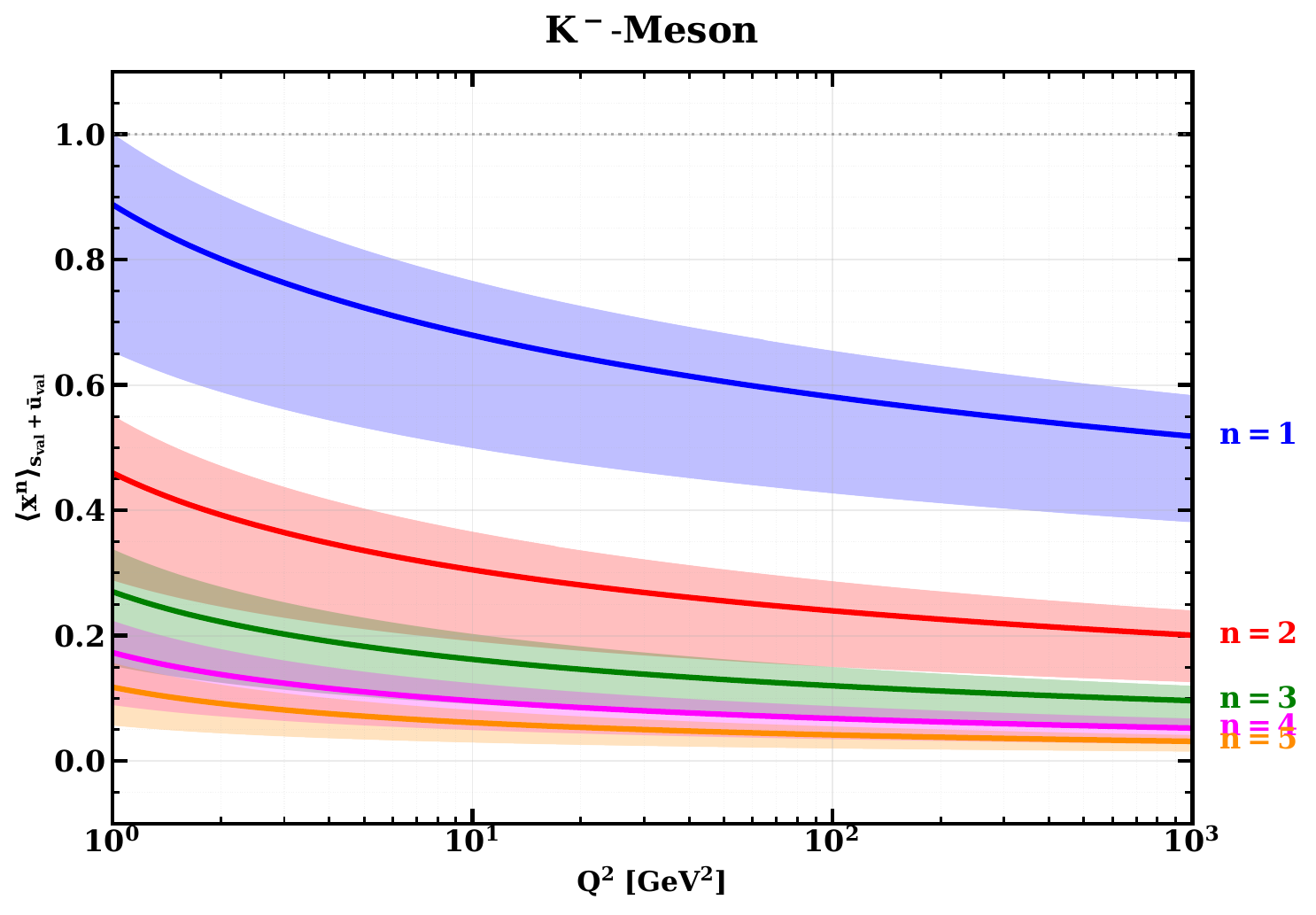}
\caption{The $n^{\rm th}$ Mellin moments of the total valence distribution, $\langle x^n\rangle_{s_{\rm val}+\bar u_{\rm val}}$, as functions of $Q^2$ for $n\le5$.}
\label{fig:kaon_mellin}
\end{subfigure}

\caption{(Color online) The PDFs of the $K^{-}(s\bar{u})$ meson evolved to $Q^2=100~\mathrm{GeV}^2$ using NLO DGLAP evolution with the initial scale varied over the range $Q_0\in[0.50,1.00]~\mathrm{GeV}$. The central curve corresponds to the initial scale $Q_0=0.75~\mathrm{GeV}$.}
\label{fig:kaon}
\end{figure*}


\begin{figure}[t]
    \centering
    \includegraphics[width=\columnwidth]{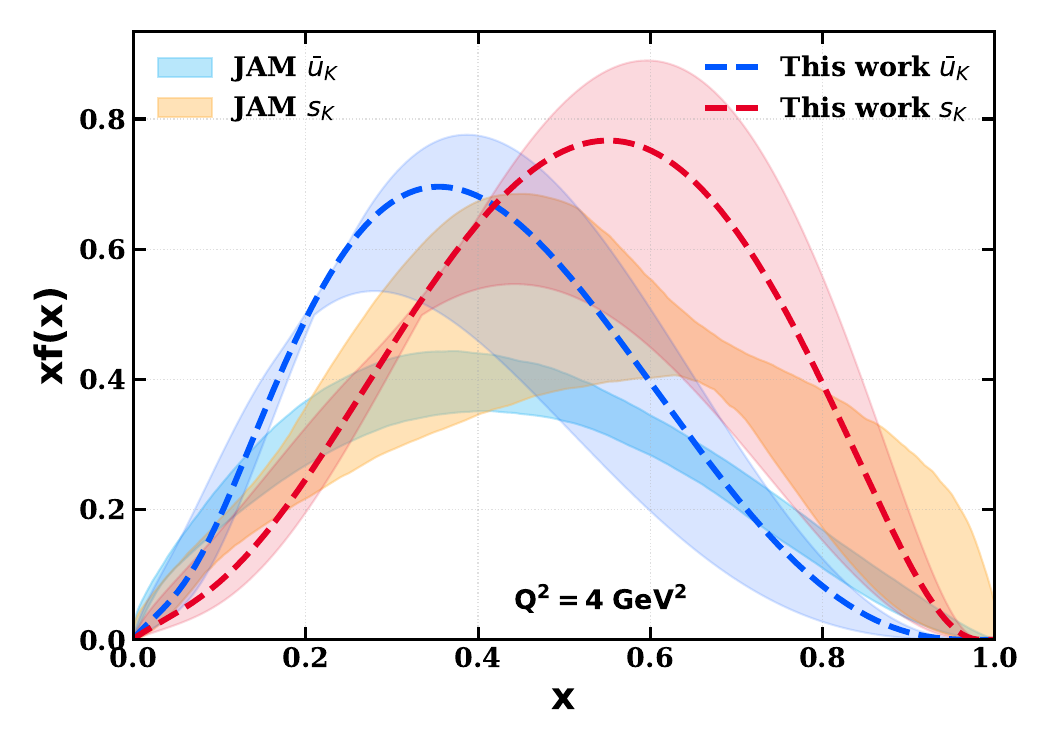}
    \caption{(Color online) Comparison of the evolved quark and antiquark PDFs of the kaon with the recent JAM analysis at $Q^2=4~\mathrm{GeV}^2$ \cite{Barry:2025wjx}. The uncertainty band corresponds to the variation of the initial scale over the range $Q_0\in[0.5,1.0]~\mathrm{GeV}$.}
    \label{kaonjam}
\end{figure}


\begin{figure*}[t]
    \centering
    \includegraphics[width=\textwidth]{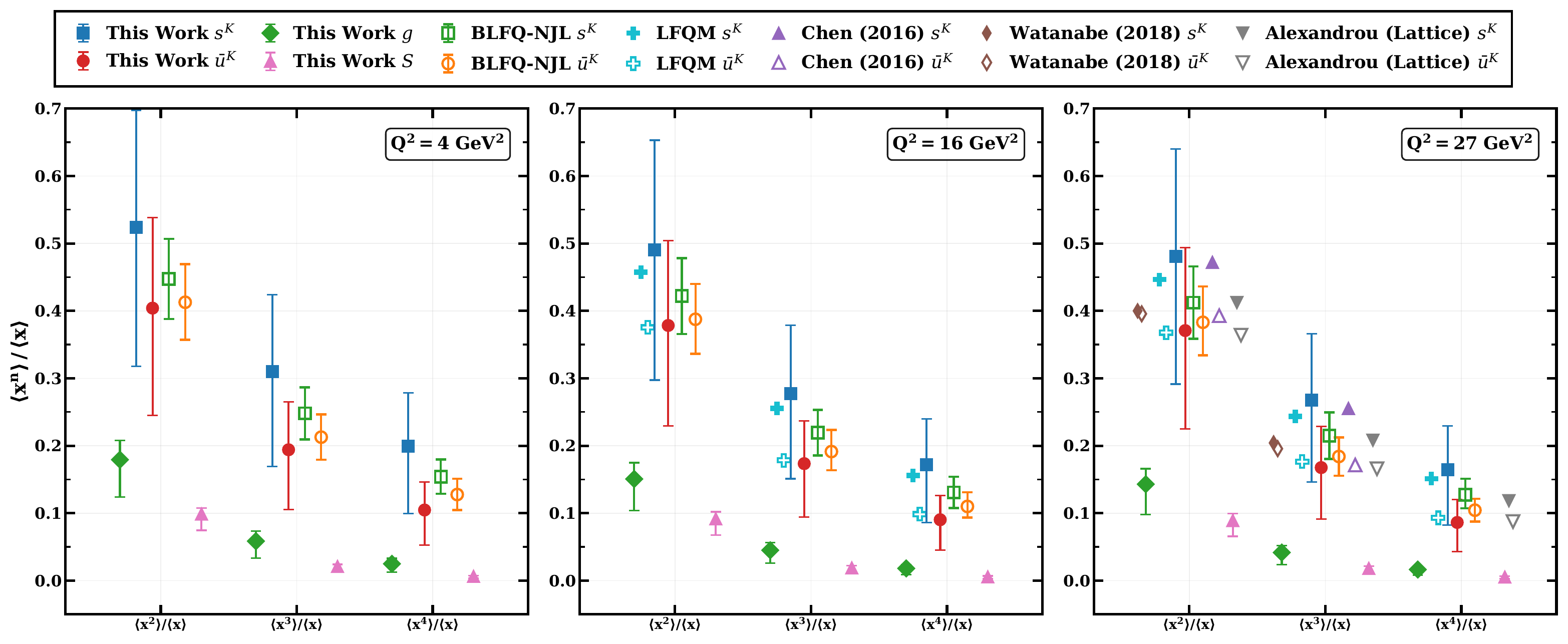}
    \caption{(Color online) Ratios of the Mellin moments, $\langle x^n\rangle/\langle x\rangle$, for the valence, gluon, and sea-quark PDFs of the kaon at $Q^2=4$, $16$, and $27~\mathrm{GeV}^2$. The results are compared with available theoretical model predictions \cite{Lan:2019rba,Choi:2025bxk,Chen:2016sno,Watanabe:2018qju} and lattice QCD results \cite{Alexandrou:2021mmi}.}
    \label{kaonMellin}
\end{figure*}


\begin{figure}[t]
    \centering
    \includegraphics[width=\columnwidth]{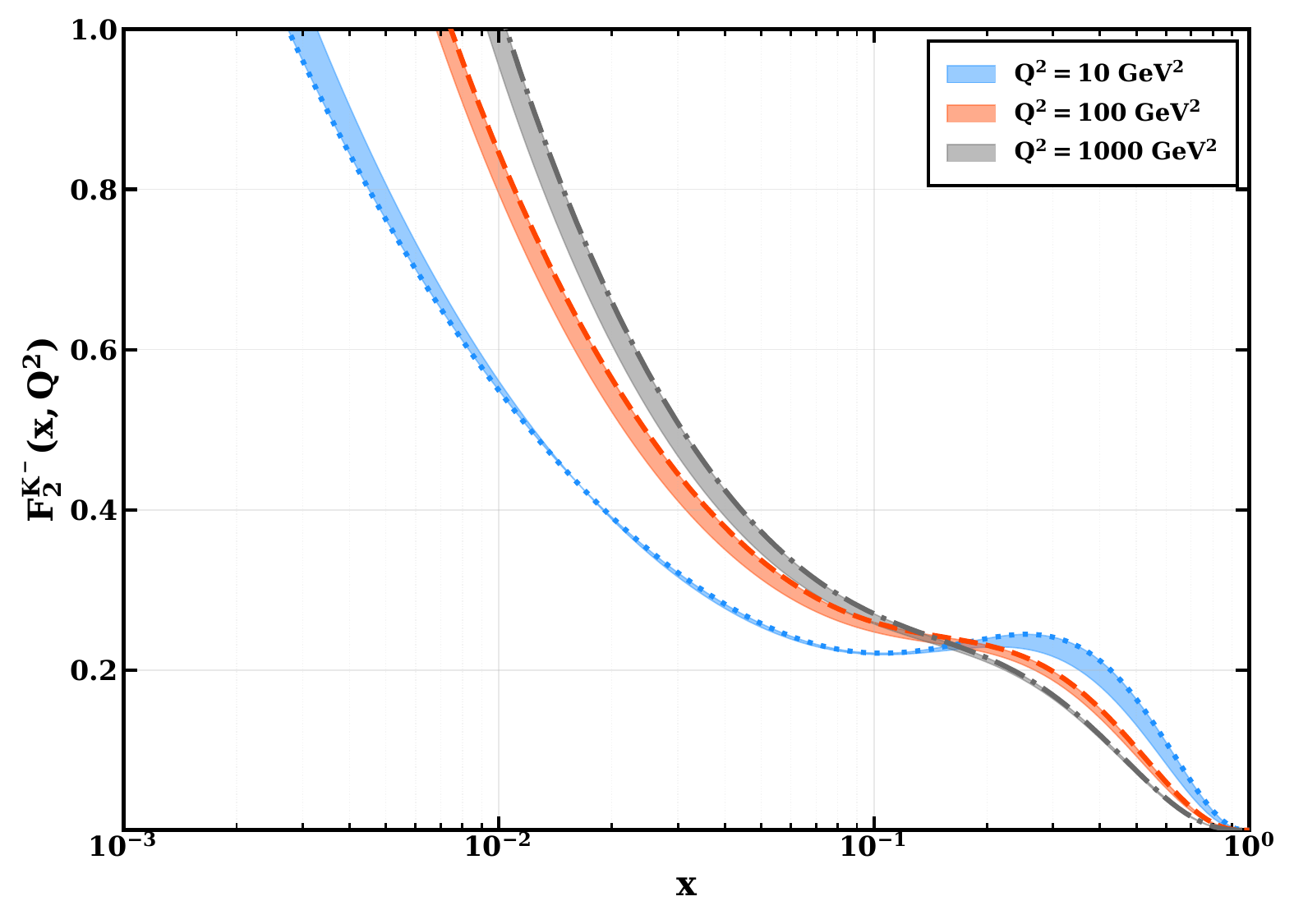}
    \caption{(Color online) The kaon structure function $F_2^{K^-}$ predicted at NLO for $Q^2=10$, $100$, and $1000~\mathrm{GeV}^2$, using the initial-scale range $Q_0\in[0.5,1.0]~\mathrm{GeV}$.}
    \label{structureF2}
\end{figure}


\begin{figure}[htbp]
    \centering
    \includegraphics[width=\linewidth]{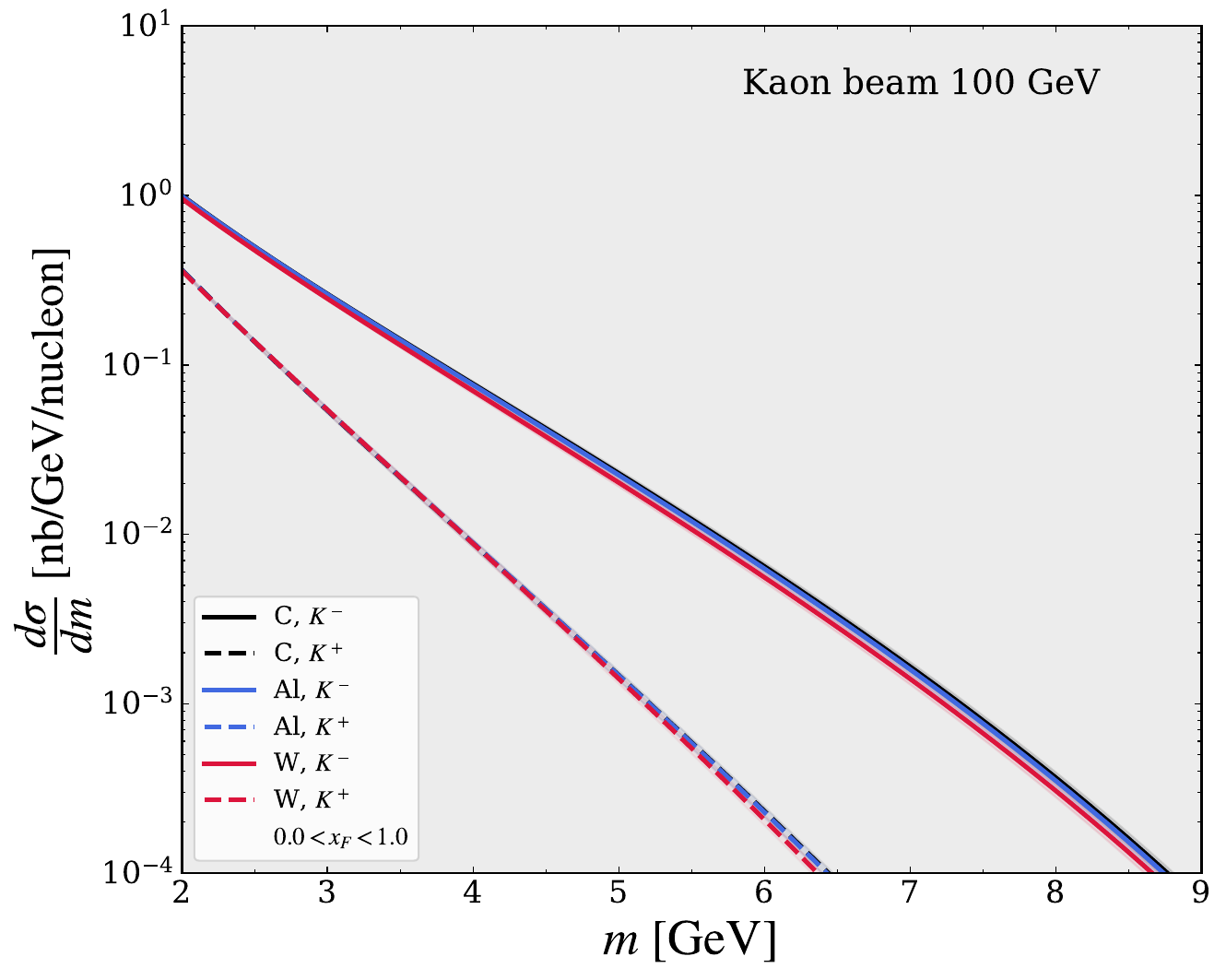}
    \caption{(Color online) The differential Drell--Yan cross section, $d\sigma/dm$, as a function of the dilepton invariant mass $m$ for $K^{+}$- and $K^{-}$-induced reactions on carbon (C), aluminum (Al), and tungsten (W) nuclear targets. The nuclear PDFs are taken from the CTEQ15 nuclear PDF set~\cite{Kovarik:2015cma}.}
    \label{dykaon}
\end{figure}


\begin{figure*}[t]
\centering

\begin{subfigure}{0.32\textwidth}
\centering
\includegraphics[width=\linewidth]{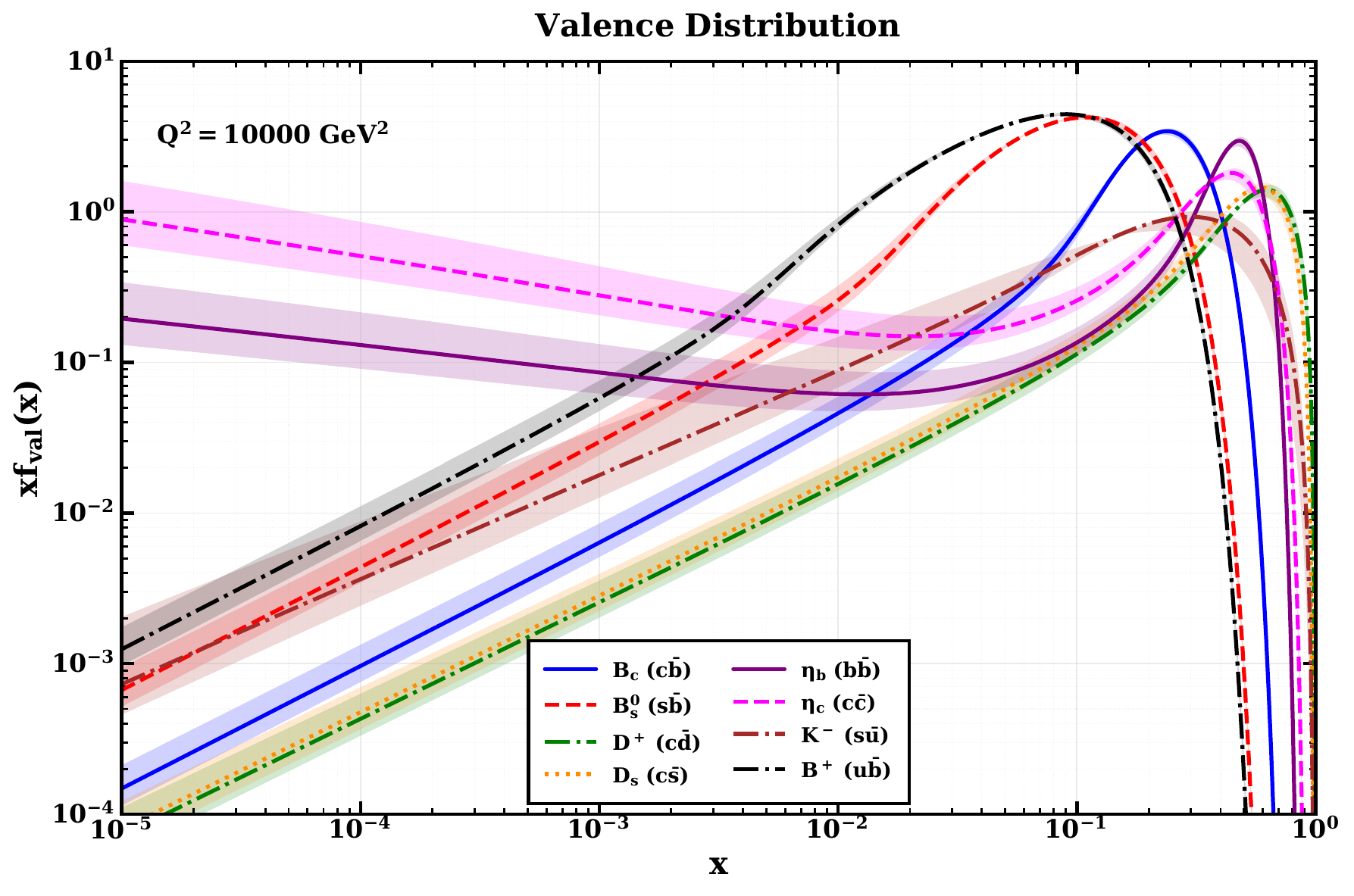}
\caption{The valence-quark distributions, $xf_{\rm val}(x)$, of different mesons evolved from their respective initial scales, shown as functions of the longitudinal momentum fraction $x$.}
\end{subfigure}
\hfill
\begin{subfigure}{0.32\textwidth}
\centering
\includegraphics[width=\linewidth]{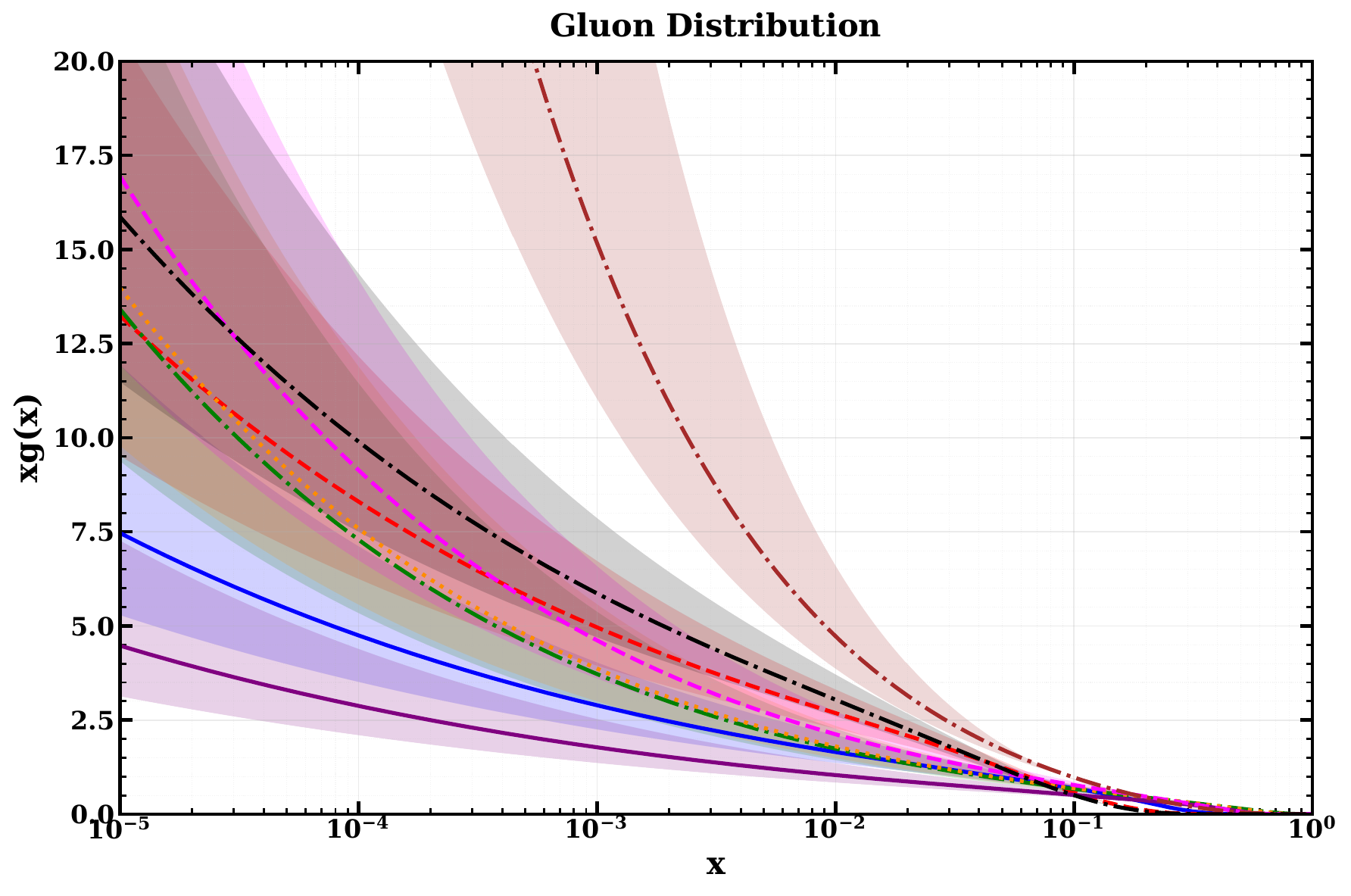}
\caption{The gluon distributions, $xg(x)$, of different mesons evolved from their respective initial scales, shown as functions of the longitudinal momentum fraction $x$.}
\end{subfigure}
\hfill
\begin{subfigure}{0.32\textwidth}
\centering
\includegraphics[width=\linewidth]{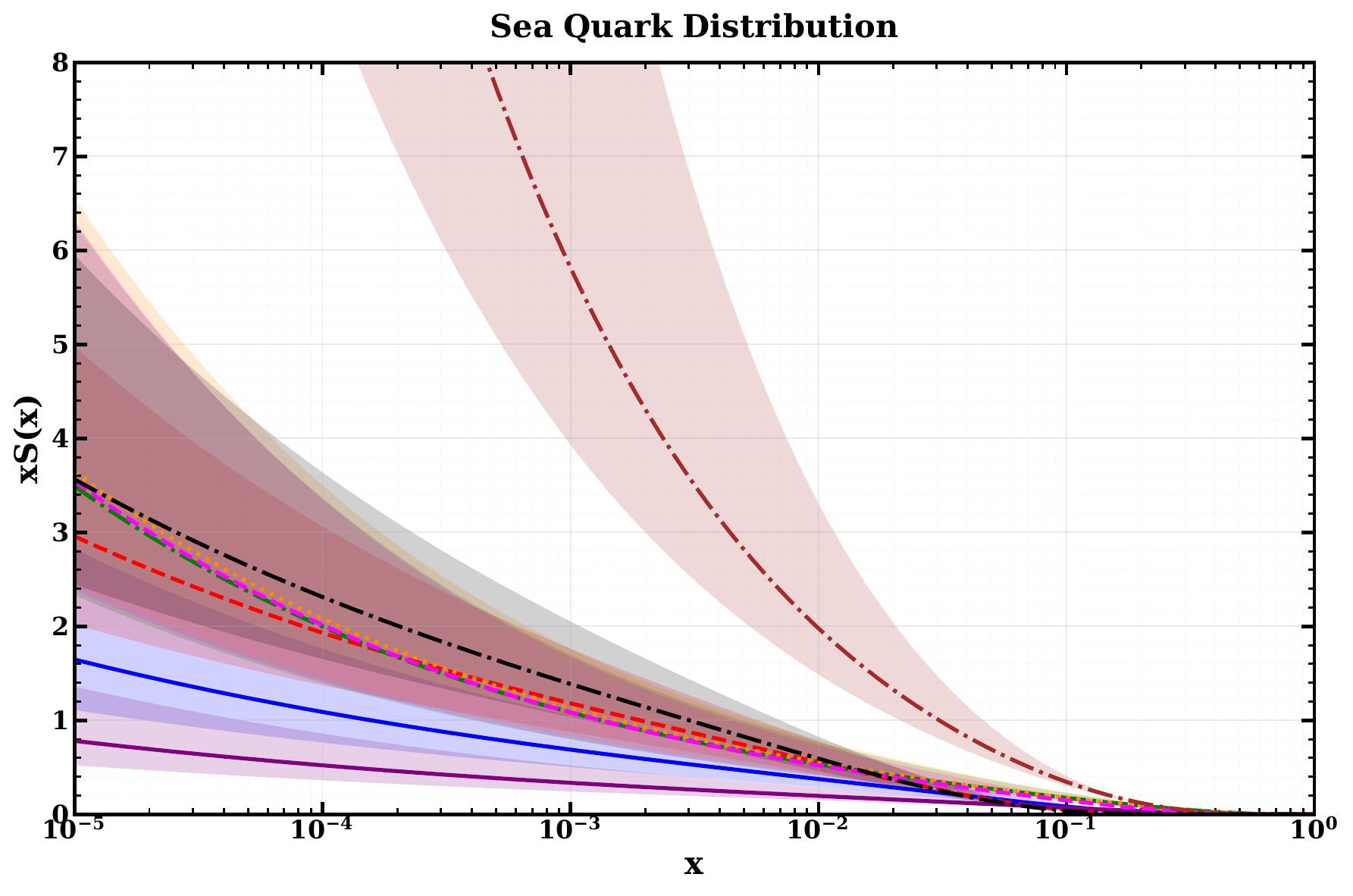}
\caption{The sea-quark distributions, $xS(x)$, of different mesons evolved from their respective initial scales, shown as functions of the longitudinal momentum fraction $x$.}
\end{subfigure}

\caption{(Color online) The constituent PDFs of different mesons evolved to $Q^2=10000~\mathrm{GeV}^2$ using their respective initial scales listed in Table~\ref{asymmetry}.}

\label{allmesons}
\end{figure*}
\begin{figure*}[t]
    \centering
    \includegraphics[width=\textwidth]{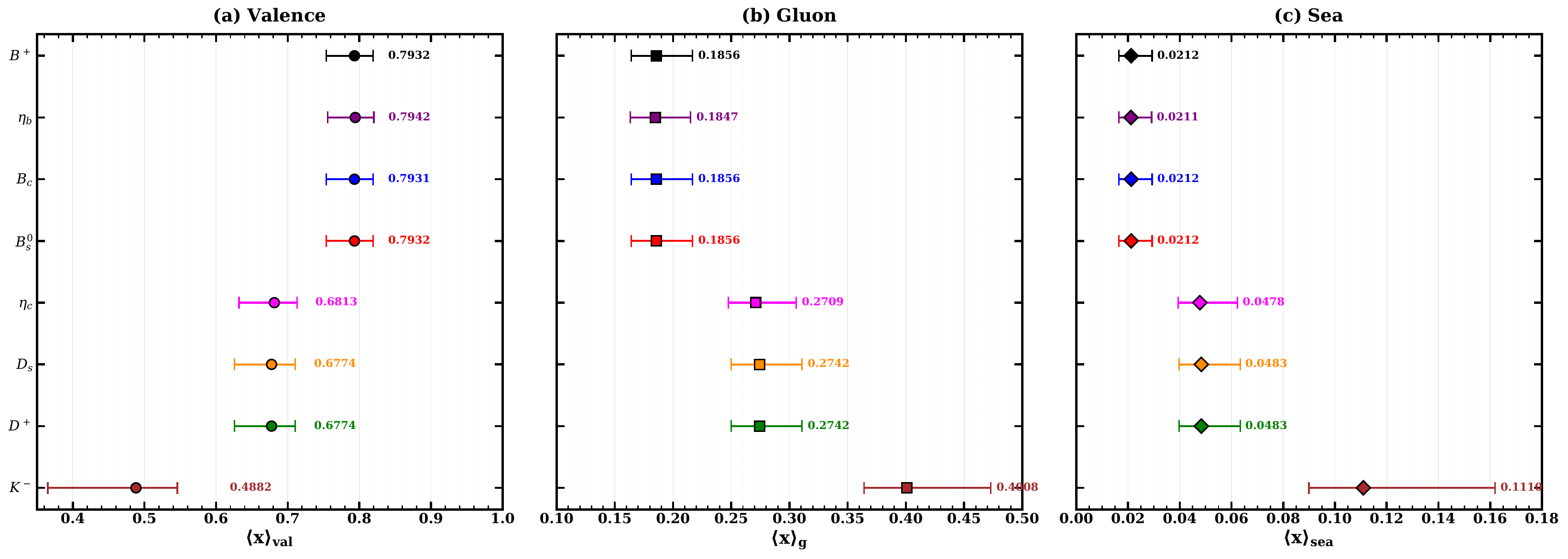}
    \caption{(Color online) Average momentum fractions carried by the total valence quark--antiquark pair, $\langle x \rangle_{\rm val}$, gluons, $\langle x \rangle_g$, and sea quarks, $\langle x \rangle_{\rm sea}$, for different mesons at $Q^2=10000~\mathrm{GeV}^2$. The error bands represent the uncertainty arising from the range of initial scales considered in this work.}
    \label{allmmvalue}
\end{figure*}

\begin{figure*}[htbp]
    \centering
    \begin{subfigure}{0.48\textwidth}
        \centering
        \includegraphics[width=\linewidth]{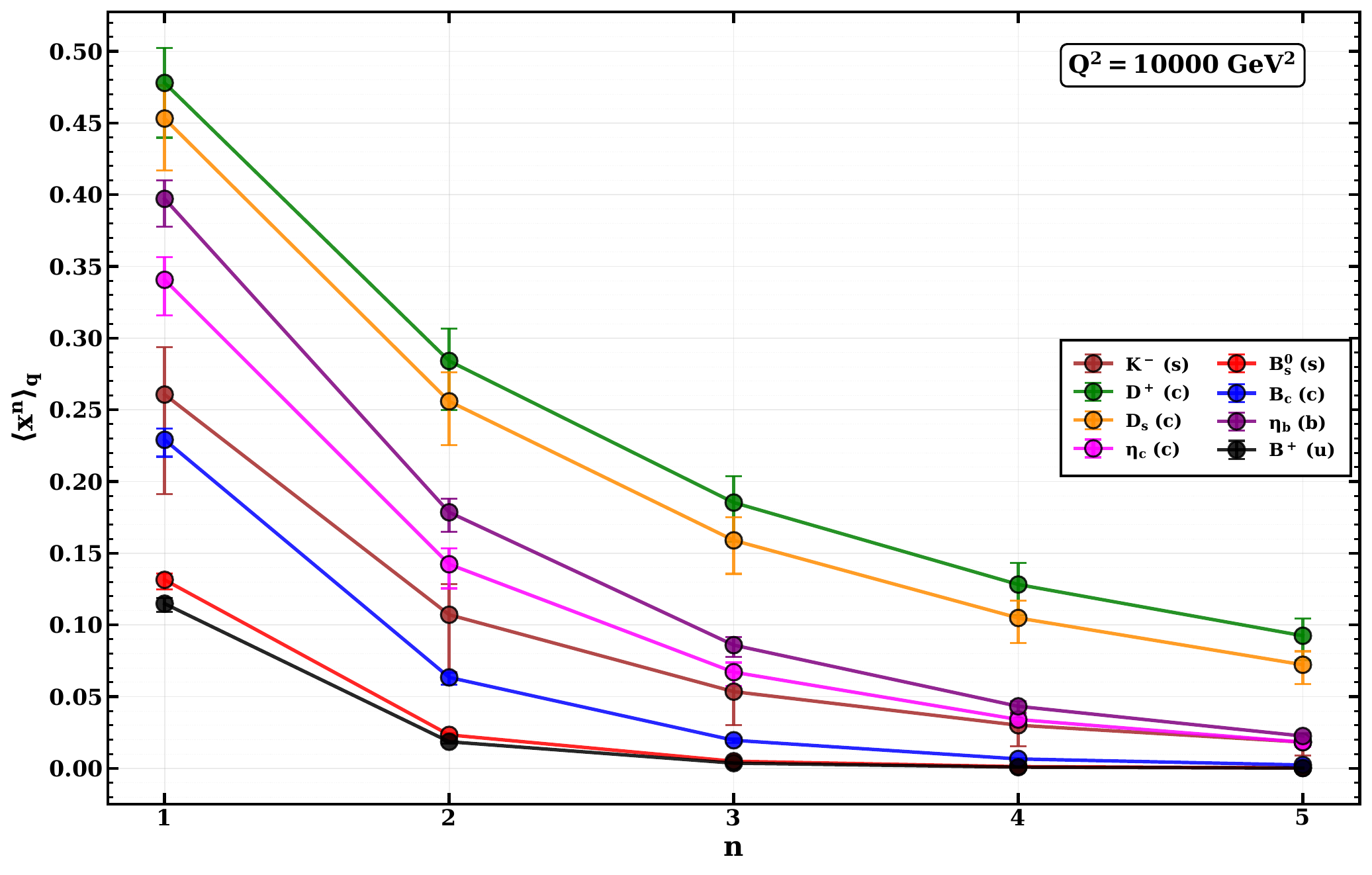}
        \caption{The Mellin moments of the valence-quark distributions for different mesons at $Q^2=10000~\mathrm{GeV}^2$ up to $n=5$.}
    \end{subfigure}
    \hfill
    \begin{subfigure}{0.48\textwidth}
        \centering
        \includegraphics[width=\linewidth]{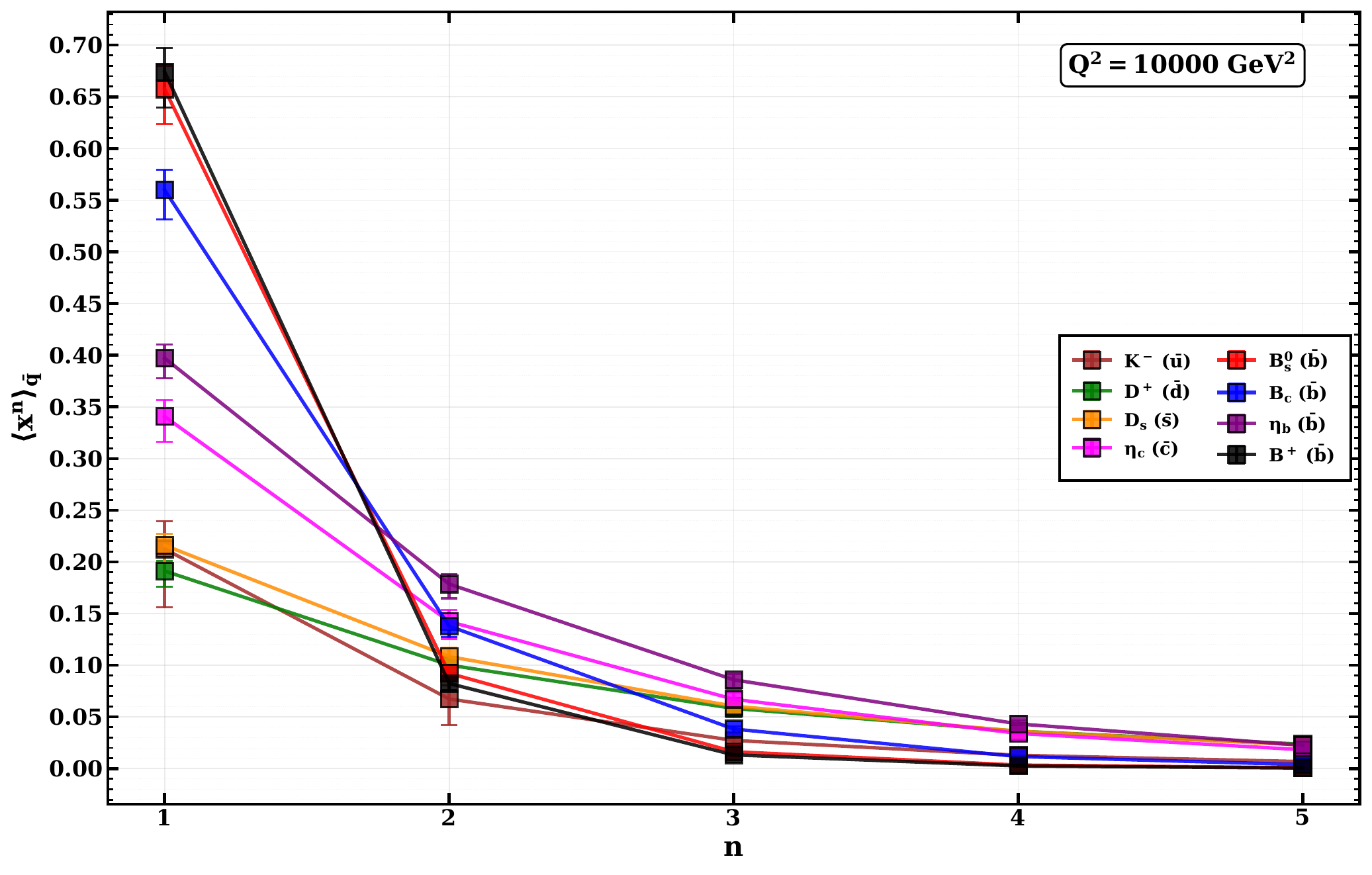}
        \caption{The Mellin moments of the valence-antiquark distributions for different mesons at $Q^2=10000~\mathrm{GeV}^2$ up to $n=5$.}
    \end{subfigure}
    \caption{(Color online) Mellin moments, $\langle x^n \rangle$, of the valence-quark and valence-antiquark distributions for various mesons evolved to $Q^2=10000~\mathrm{GeV}^2$. The error bands represent the uncertainty associated with varying the initial scale $Q_0$ within the range considered in this work.}
    \label{fig:Mellin_momentsall}
\end{figure*}

\begin{figure}[t]
    \centering
    \includegraphics[width=\columnwidth]{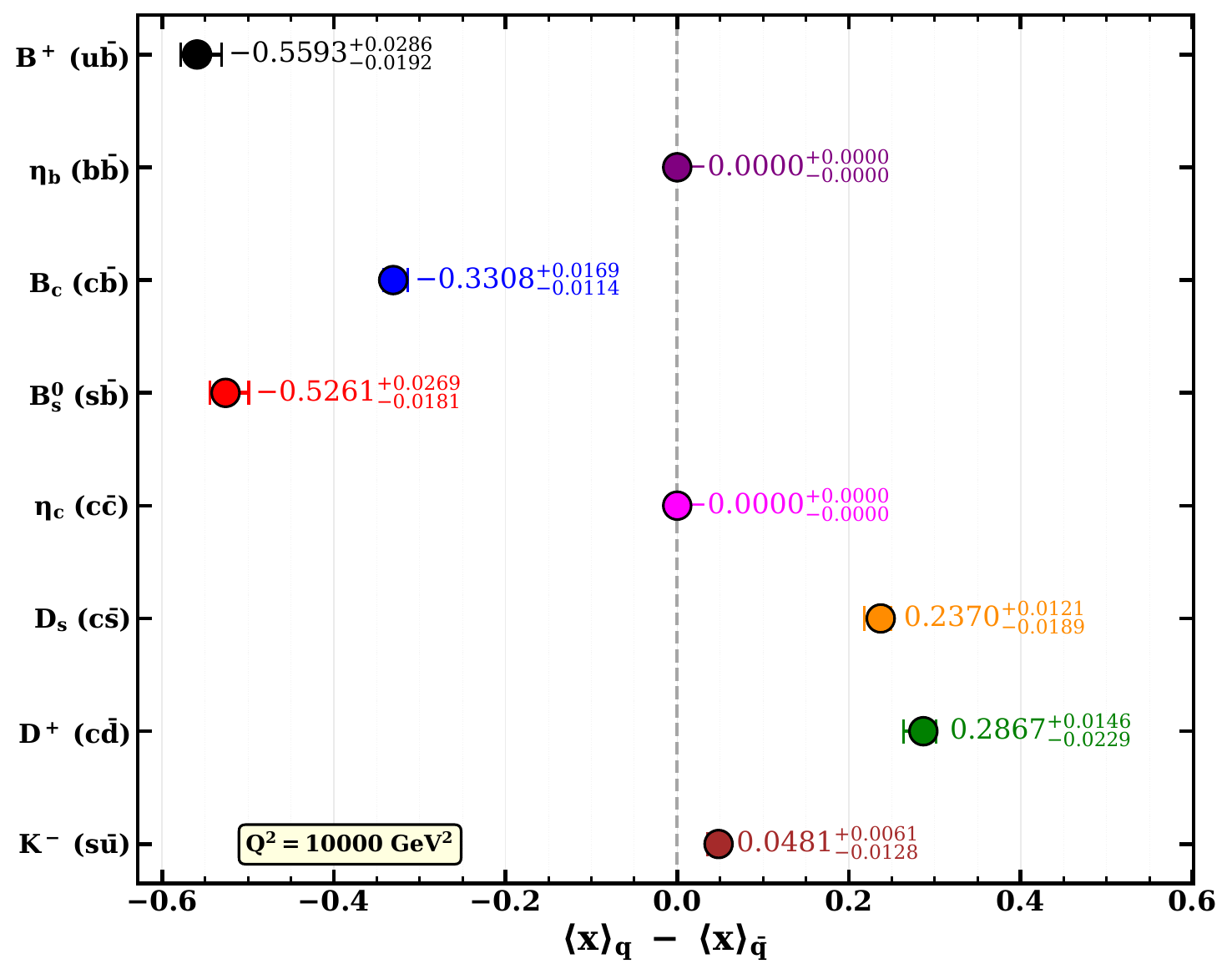}
    \caption{(Color online) Difference between the average momentum fractions carried by the valence quark and valence antiquark, $\langle x \rangle_q-\langle x \rangle_{\bar q}$, for different mesons at $Q^2=10000~\mathrm{GeV}^2$. The error bands represent the uncertainty associated with varying the initial scale $Q_0$ within the range considered in this work.}
    \label{fixmelli}
\end{figure}


\begin{figure}[t]
    \centering
    \includegraphics[width=\columnwidth]{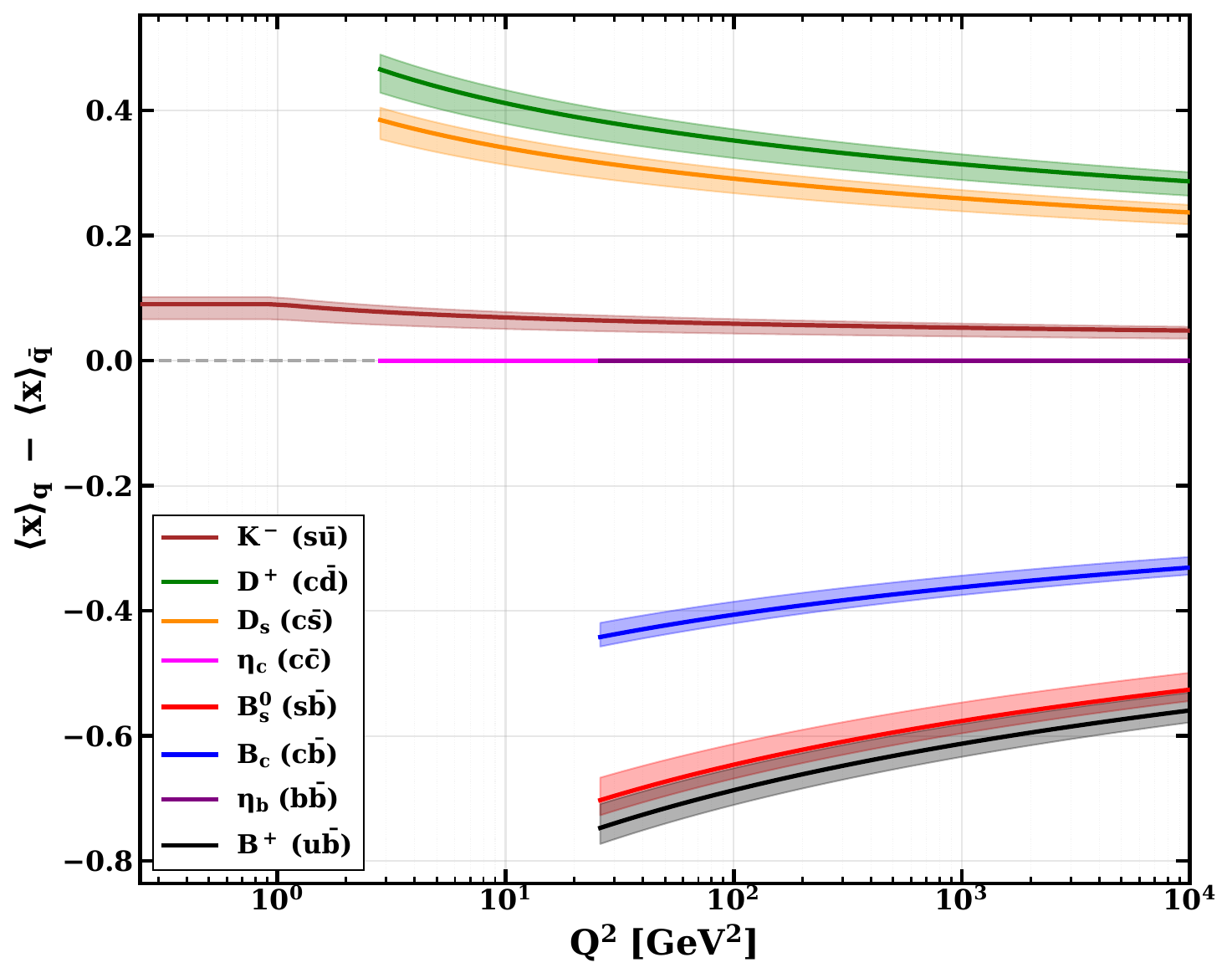}
    \caption{(Color online) Difference between the average momentum fractions carried by the valence quark and valence antiquark, $\langle x \rangle_q-\langle x \rangle_{\bar q}$, as a function of $Q^2$. The error bands represent the uncertainty associated with varying the initial scale $Q_0$ within the range considered in this work.}
    \label{differenceMellin}
\end{figure}


\begin{figure}[t]
    \centering
    \includegraphics[width=\columnwidth]{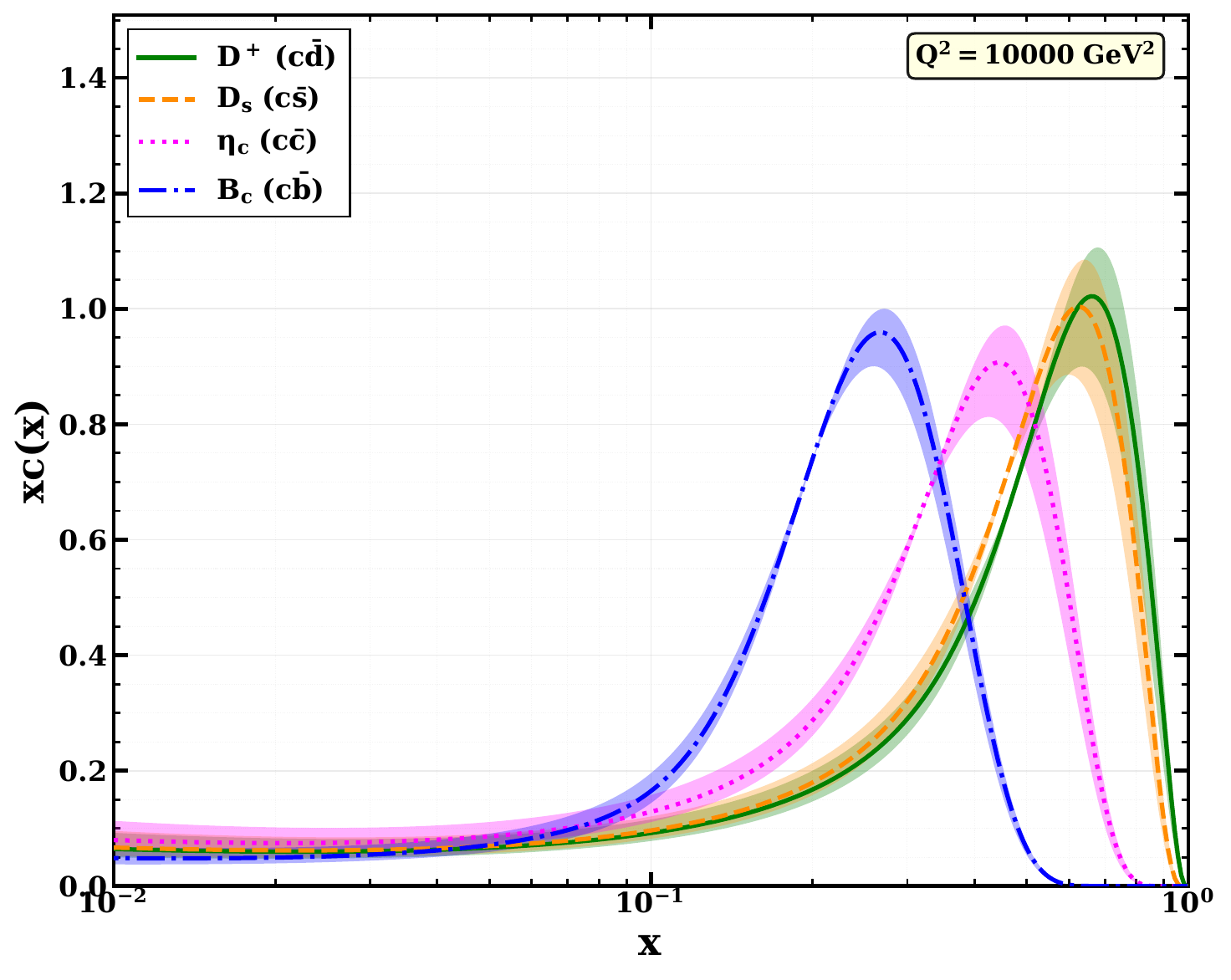}
    \caption{(Color online) Charm-quark distribution, $xc(x)$, as a function of the longitudinal momentum fraction $x$ at $Q^{2}=10^{4}~\mathrm{GeV}^{2}$ for the $D^{+}(c\bar d)$, $D_s^{+}(c\bar s)$, $\eta_c(c\bar c)$, and $B_c^{+}(c\bar b)$ mesons. The error bands represent the uncertainty associated with varying the initial scale $Q_0$ within the range considered in this work.}
    \label{cquark}
\end{figure}


\begin{figure}[t]
    \centering
    \includegraphics[width=\columnwidth]{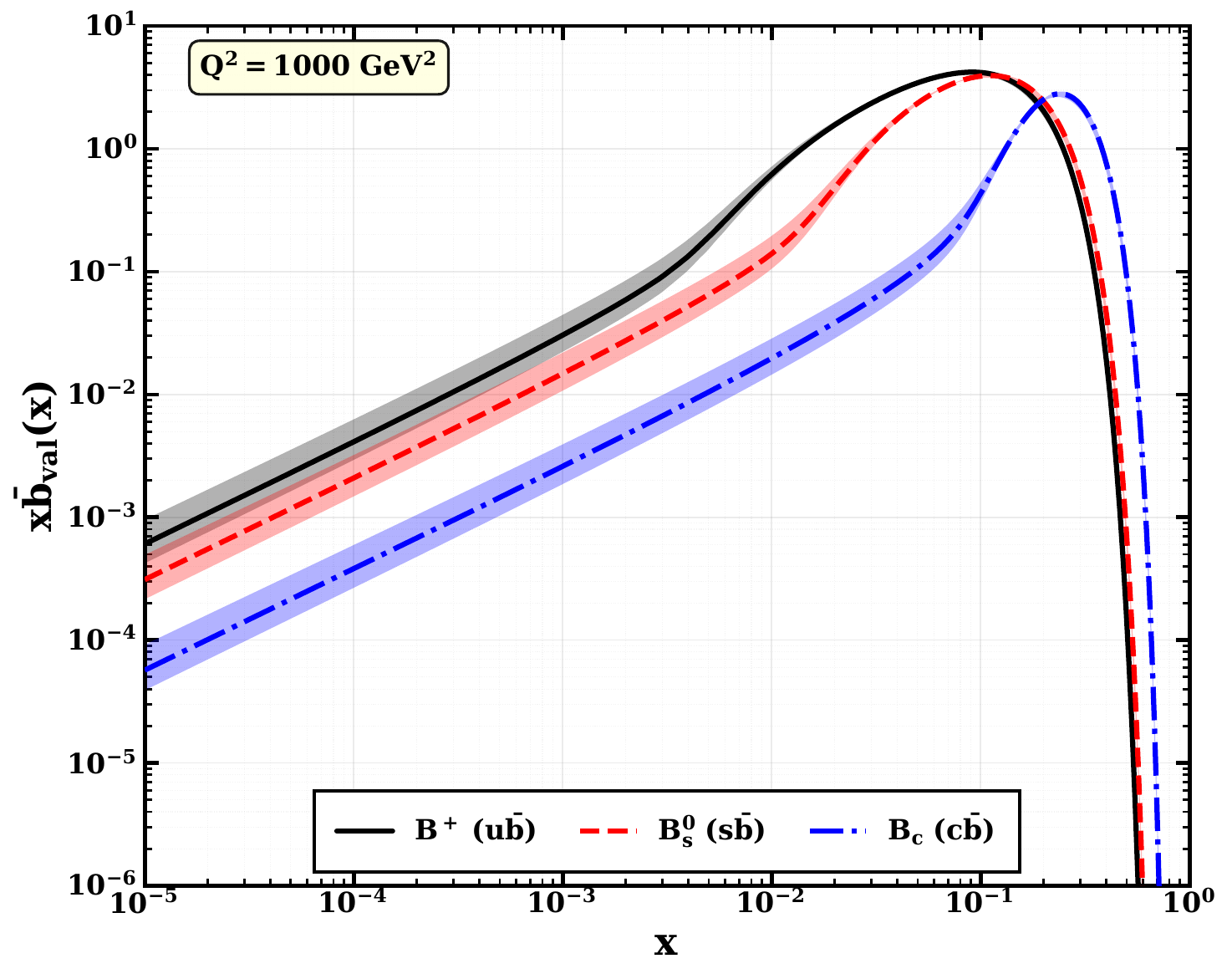}
    \caption{(Color online) Bottom-quark distribution, $xb(x)$, as a function of the longitudinal momentum fraction $x$ at $Q^{2}=10^{4}~\mathrm{GeV}^{2}$ for the $B^{+}(u\bar b)$, $B_s^{0}(s\bar b)$, $B_c^{+}(c\bar b)$, and $\eta_b(b\bar b)$ mesons. The error bands represent the uncertainty associated with varying the initial scale $Q_0$ within the range considered in this work.}
    \label{bquark}
\end{figure}


\begin{figure}[t]
    \centering
    \includegraphics[width=\columnwidth]{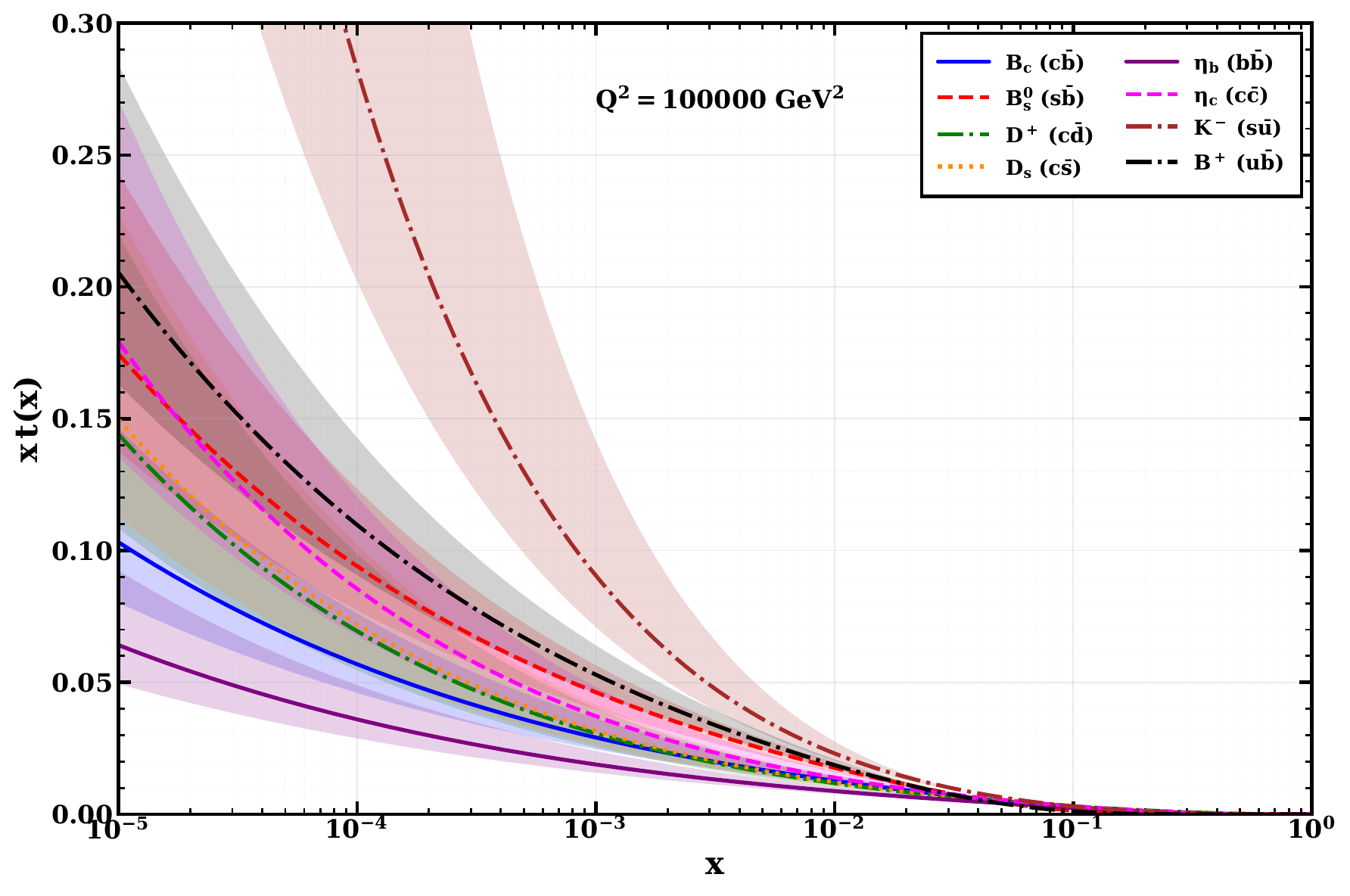}
    \caption{(Color online) Top sea-quark distributions, $xt(x)$, for different mesons at $Q^{2}=10^{5}~\mathrm{GeV}^{2}$. The error bands represent the uncertainty associated with varying the initial scale $Q_0$ within the range considered in this work.}
    \label{top}
\end{figure}


\begin{table}[t]
\centering
\caption{Ranges of the initial scale, $Q_0$, used for different mesons in this work.}
\setlength{\tabcolsep}{4pt}
\renewcommand{\arraystretch}{1.0}
\begin{tabular}{cc}
\hline
Meson & $Q_0$ (GeV) \\
\hline
$K^{-}$ & $0.50$--$1.00$ \\

$D^{+},\,D_s,\,\eta_c$ & $1.68$--$3.36$ \\

$B_s^{0},\,B_c,\,\eta_b,\,B^{+}$ & $5.10$--$10.20$ \\

\hline
\end{tabular}
\label{asymmetry}
\end{table}
The probability of finding valence quarks, antiquarks, gluons, and sea quarks inside a meson as functions of the longitudinal momentum fraction is encoded in the PDFs. For pseudoscalar mesons, there exists only a single leading-twist unpolarized quark PDF, in contrast to three for spin-$\frac{1}{2}$ nucleons and four for spin-1 mesons. This unpolarized PDF, $f_q(x)$, arises from the non-flip helicity configurations of the quark--antiquark pair within an unpolarized meson. At a fixed light-front time $z^+=\tau$, the quark--quark correlator defining the unpolarized quark PDF is given by \cite{P:2026crg,Maji:2016yqo}
\begin{equation}
\begin{aligned}
f_q(x)
&=
\frac{1}{2}
\int \frac{dz^-}{4\pi}\,
e^{ i k^{+} z^{-}/2}
\Big\langle \Psi_h(P^+,\mathbf{P}_\perp,S_z=0)\Big|
\\
&\qquad
\bar{\psi}(0)\, \mathcal{W}(0,z)\,\Gamma\,\psi(z)
\Big|\Psi_h(P^+,\mathbf{P}_\perp,S_z=0)\Big\rangle
\Big|_{z^{+},\mathbf{z}_{\perp}=0}.
\end{aligned}
\end{equation}

Here, $\Gamma=\gamma^+$ denotes the LF vector current corresponding to the unpolarized quark PDF, which determines the Lorentz structure of the correlator. The quantity $\psi(z)$ is the quark field operator, and $z=(z^+,z^-,z_\perp)$ denotes the space-time four-vector specifying the separation between the quark field operators. The Wilson line $\mathcal{W}(0,z)$ ensures the gauge invariance of the bilocal quark-field operators appearing in the correlation function \cite{Bacchetta:2020vty}. In the present work, the Wilson line is taken to be unity.

\par Using the meson Fock state in Eq.~(\ref{eqeq}) together with the quark field operators given in Ref.~\cite{Harindranath:1996hq}, the overlap representation of the unpolarized quark PDF, $f_q(x)$, can be written as
\begin{equation}
\begin{aligned}
f_q(x)
&=
\int \frac{d^{2}\mathbf{k}_\perp}{16\pi^{3}}
\, |\phi(x,\mathbf{k}_\perp^{2})|^{2}
\\
&\times
\Big[
|\mathcal{S}(x,\mathbf{k}_\perp,\uparrow,\uparrow)|^{2}
+
|\mathcal{S}(x,\mathbf{k}_\perp,\downarrow,\downarrow)|^{2}
\\
&+
|\mathcal{S}(x,\mathbf{k}_\perp,\downarrow,\uparrow)|^{2}
+
|\mathcal{S}(x,\mathbf{k}_\perp,\uparrow,\downarrow)|^{2}
\Big].
\end{aligned}
\end{equation}

Substituting the spin wave functions given in Eq.~(\ref{spin1}), the explicit expression for the quark PDF becomes
\begin{equation}
\begin{split}
f_q(x) = &\int \frac{d^{2}\mathbf{k}_\perp}{16\pi^{3}}
\, |\phi(x,\mathbf{k}_\perp^{2})\, \mathcal{A}_{q \bar q}|^2
\\
&\times
\frac{\textbf{k}_\perp^2+\Big((1-x)m_q+x m_{\bar q}\Big)^2}
{\omega^2}.
\end{split}
\end{equation}
One can obtain the antiquark PDF, $f_{\bar q}(x)$, from the quark PDF using the momentum sum rule,
$f_{\bar q}(x)=f_q(1-x)$.
This relation implies that, at the initial scale, the total longitudinal momentum of the parent meson is entirely carried by the constituent quark and antiquark. The unpolarized quark and antiquark PDFs satisfy the following sum rules \cite{Puhan:2023ekt,Puhan:2023hio}:
\begin{equation}
\begin{alignedat}{2}
&\int_{0}^{1} dx\, f_q(x,Q_0^2) = 1 , \\
&\int_{0}^{1} dx\, f_{\bar q}(x,Q_0^2) = 1 , \\
&\int_{0}^{1} dx\, x\,\big[f_q(x,Q_0^2)
+f_{\bar q}(x,Q_0^2)\big] &&= 1 .
\end{alignedat}
\label{eq19}
\end{equation}

Since gluon and sea-quark contributions are not included at the initial scale in the present work, their corresponding momentum fractions vanish:
\begin{equation}
\begin{aligned}
\int_{0}^{1} dx\, g(x,Q_0^2) &= 0 , \\
\int_{0}^{1} dx\, S(x,Q_0^2) &= 0 .
\end{aligned}
\end{equation}
Here, $g(x,Q_0^2)$ and $S(x,Q_0^2)$ denote the gluon and sea-quark distributions, respectively. It should be noted that $Q_0$ refers to the lower limit of the initial-scale range considered in this work. The dynamical gluon contributions arising from higher Fock states in the pion and kaon are currently under investigation, and the corresponding results will be presented in a future publication.

\subsection{Initial Scale PDFs}
For the present study, we consider the ground-state pseudoscalar mesons investigated in our previous work \cite{Puhan:2024jaw}, namely the kaon $K^-(s\bar{u})$, the $D$ mesons $\left(D^+(c\bar{d}),\,D_s^+(c\bar{s})\right)$, the charmonium state $\eta_c(c\bar{c})$, the bottomonium state $\eta_b(b\bar{b})$, and the $B$ mesons $\left(B^+(u\bar{b}),\,B_c^+(c\bar{b}),\,B_s^0(s\bar{b})\right)$. A detailed study of the pion PDFs within the LCQM has already been presented in our previous work \cite{P:2026crg}. Among these mesons, the kaon is of particular experimental interest since its PDFs can be accessed through the ongoing COMPASS++/AMBER kaon-induced Drell--Yan program at CERN, as well as future measurements at the EIC. In contrast, the PDFs of the heavier mesons are difficult to probe experimentally because of their very short lifetimes. Consequently, theoretical studies provide an important avenue for understanding their partonic structure.

The numerical calculations require only two model parameters: the constituent quark masses, $m_{q(\bar q)}$, and the harmonic scale parameter, $\beta_h$, whose values are listed in Table~\ref{tab1}. The resulting quark PDFs at the initial scale are shown in Fig.~\ref{allpdf}. All distributions are positive over the entire kinematic region, while the antiquark PDFs are related to the quark PDFs through the symmetry relation discussed in Sec.~\ref{pdf}. Mesons composed of equal-mass constituents, namely $\eta_c$ and $\eta_b$, exhibit symmetric PDFs about $x=0.5$. In contrast, for heavy--light systems such as the kaon and the $D$ mesons, the distributions peak at $x>0.5$, indicating that the heavier constituent carries a larger fraction of the longitudinal momentum. The opposite behavior is observed for the $B$ mesons, where the heavy antiquark carries most of the meson momentum, leading to quark distributions that peak at $x<0.5$.

The constituent-quark mass also has a significant influence on the shape of the PDFs. As the meson mass increases, the distributions become increasingly localized in $x$, resulting in narrower distributions with larger peak values. This trend is clearly illustrated by the comparison between the $\eta_c$ and $\eta_b$ PDFs, where the heavier $\eta_b$ exhibits a noticeably narrower distribution. Among all the mesons considered, the $B^+$ meson possesses the most localized valence distribution. Physically, this behavior reflects the increasingly non-relativistic nature of heavy-quark systems. As the constituent mass increases, the quark and antiquark are more likely to share the meson momentum almost equally, suppressing the probability of finding constituents carrying very small longitudinal momentum fractions. Similar features have also been reported within the basis light-front quantization (BLFQ) framework \cite{Lan:2019img}, Bethe--Salpeter wave functions (BSWFs) \cite{Shi:2021nvg,Shi:2022erw}, the light-front quark model (LFQM) \cite{Acharyya:2024tql,Wu:2025rto}, the holographic basis approach \cite{Li:2015zda,Li:2017mlw,Tang:2018myz,Tang:2019gvn}, and Ref.~\cite{Almeida-Zamora:2026qnj}.
\begin{table*}[t]
\caption{
First and second Mellin moments of the valence quark, antiquark, gluon and sea-quarks $K^-$-meson
at $Q^2=4$, $16$ and $27$ GeV$^2$ scales.
The uncertainties correspond to the variation of the initial scale
$Q_0\in[0.50,1.00]~\mathrm{GeV}$.}
\label{kaon_moments}
\centering
\renewcommand{\arraystretch}{1.15}
\setlength{\tabcolsep}{5pt}
\begin{ruledtabular}
\begin{tabular}{cccccc}
$Q^2~(\mathrm{GeV}^2)$
& Moment
& $\langle x^n\rangle_{s_{\mathrm{val}}}$
& $\langle x^n\rangle_{\bar{u}_{\mathrm{val}}}$
& $\langle x^n\rangle_{g}$
& $\langle x^n\rangle_{\mathrm{sea}}$
\\
4
& $\langle x\rangle$
& $0.428^{+0.054}_{-0.114}$
& $0.312^{+0.039}_{-0.083}$
& $0.236^{+0.130}_{-0.209}$
& $0.024^{+0.037}_{-0.012}$
\\

& $\langle x^2\rangle$
& $0.231^{+0.046}_{-0.086}$
& $0.133^{+0.026}_{-0.049}$
& $0.043^{+0.006}_{-0.010}$
& $0.0025^{+0.0023}_{-0.0011}$
\\
\hline
16
& $\langle x\rangle$
& $0.378^{+0.048}_{-0.101}$
& $0.276^{+0.035}_{-0.073}$
& $0.305^{+0.124}_{-0.189}$
& $0.041^{+0.041}_{-0.015}$
\\

& $\langle x^2\rangle$
& $0.192^{+0.038}_{-0.071}$
& $0.110^{+0.022}_{-0.041}$
& $0.046^{+0.001}_{-0.004}$
& $0.0039^{+0.0019}_{-0.0011}$
\\
\hline
27
& $\langle x\rangle$
& $0.364^{+0.046}_{-0.097}$
& $0.266^{+0.034}_{-0.071}$
& $0.322^{+0.122}_{-0.184}$
& $0.048^{+0.042}_{-0.016}$
\\

& $\langle x^2\rangle$
& $0.181^{+0.036}_{-0.067}$
& $0.104^{+0.021}_{-0.039}$
& $0.046^{+0.000}_{-0.003}$
& $0.0045^{+0.0017}_{-0.0011}$
\\
\end{tabular}
\end{ruledtabular}
\end{table*}
\par We have also calculated the Mellin moments, $\langle x^n\rangle$, of the quark and antiquark PDFs for different mesons, which are defined as
\begin{eqnarray}
\langle x^n \rangle=
\frac{\int dx\, x^{n} f(x,Q^2)}
{\int dx\, f(x,Q^2)}.
\end{eqnarray}
The Mellin moments at the initial scale $Q_0$ are listed in Table~\ref{tab2}. The zeroth Mellin moment corresponds to the valence quark number and is equal to unity for each constituent, satisfying the quark number sum rule. The first Mellin moment gives the average longitudinal momentum fraction carried by the constituent quark or antiquark. At the model scale, the total momentum is shared entirely by the valence quark and antiquark, consistent with the momentum sum rule in Eq.~(\ref{eq19}).

The calculated Mellin moments reveal a clear correlation between the constituent mass and the average longitudinal momentum fraction. Within a given meson, the heavier constituent always carries a larger fraction of the meson momentum, which can be approximately expressed as
\begin{eqnarray}
\langle x\rangle_{q(\bar q)}
\propto
m_{q(\bar q)}.
\end{eqnarray}
Consequently, the bottom antiquark in the $B$-meson family carries the largest momentum fraction among all constituents considered in this work, whereas mesons composed of equal-mass constituents, such as $\eta_c$ and $\eta_b$, exhibit identical quark and antiquark Mellin moments. The momentum asymmetry,
$\langle x\rangle_q-\langle x\rangle_{\bar q}$,
is therefore governed primarily by the constituent mass difference and becomes most pronounced for the heavy--light $B$ mesons.

The Mellin moments further demonstrate the suppression of the longitudinal momentum carried by a light constituent in the presence of a heavier partner. For a given quark flavor, the Mellin moments at the initial scale satisfy
\begin{equation}
\begin{aligned}
\langle x^n \rangle_u^{K^-}
&>
\langle x^n \rangle_u^{B^+},\\
\langle x^n \rangle_s^{K^-}
&>
\langle x^n \rangle_s^{D_s}
>
\langle x^n \rangle_s^{B_s^0},\\
\langle x^n \rangle_c^{D^+}
&>
\langle x^n \rangle_c^{D_s}
>
\langle x^n \rangle_c^{\eta_c}
>
\langle x^n \rangle_c^{B_c^+},\\
\langle x^n \rangle_b^{B^+}
&>
\langle x^n \rangle_b^{B_s^0}
>
\langle x^n \rangle_b^{B_c^+}
>
\langle x^n \rangle_b^{\eta_b},
\label{ruleee}
\end{aligned}
\end{equation}
showing that the momentum carried by a given quark flavor decreases as the mass of its partner constituent increases.

Although the model provides the non-perturbative PDFs at the initial scale, phenomenological applications require their evolution to higher momentum scales. Therefore, throughout this work the model PDFs are evolved using the standard DGLAP evolution equations at NLO accuracy with the \textsc{HOPPET} toolkit \cite{Salam:2008qg}, thereby connecting the non-perturbative model scale with the perturbative regime relevant for experimental measurements.
\par The DGLAP evolution equations relate the PDFs at different energy scales and provide the connection between the non-perturbative model scale and the perturbative  \cite{Gribov:1972ri,Lipatov:1974qm,Altarelli:1977zs,Dokshitzer:1977sg}
\begin{align}
\frac{\partial f(x,Q^{2})}{\partial \ln Q^{2}}
&=
\frac{\alpha_s(Q^{2})}{2\pi}
\left[
P_{qq}\otimes q + P_{qg}\otimes g
\right],
\\
\frac{\partial g(x,Q^{2})}{\partial \ln Q^{2}}
&=
\frac{\alpha_s(Q^{2})}{2\pi}
\left[
P_{gq}\otimes q + P_{gg}\otimes g
\right],
\end{align}
where
\begin{equation}
(P \otimes f)(x)=\int_x^1 \frac{dz}{z}\,
P\!\left(\frac{x}{z}\right)f(z,Q^{2}).
\end{equation}
Here, $\alpha_s(Q^{2})$ denotes the strong coupling constant, which has been taken from the CT14 proton PDF set available in the LHAPDF library \cite{Buckley:2014ana}. The variable $z$ represents the longitudinal momentum fraction of the parent parton before splitting, while $P_{qq}$, $P_{qg}$, $P_{gq}$, and $P_{gg}$ are the standard DGLAP splitting kernels. Starting from the valence-quark distributions obtained within the LCQM, QCD evolution dynamically generates gluon and sea-quark distributions through successive parton splittings, thereby providing the complete partonic structure of the mesons at higher momentum scales.

\subsection{Evolved Constituent PDFs}
The quark and antiquark PDFs shown in Fig.~\ref{allpdf} are evolved to higher energy scales using the NLO DGLAP evolution equations. For each meson, the initial scale is varied within the range $Q_0$--$2Q_0$, with the corresponding values listed in Table~\ref{asymmetry}. The lower limit of the evolution scale is chosen such that $Q_0\ge m_Q$, where $m_Q$ denotes the mass of the heaviest constituent quark in the corresponding meson. This choice provides a physically motivated starting point for perturbative evolution, where the hadronic structure is still dominated by the valence quark-antiquark configuration while remaining above the heavy-quark threshold.

Among all the mesons considered in this work, the kaon is of particular phenomenological interest because its PDFs can be directly constrained by the ongoing COMPASS++/AMBER program and future EIC measurements. The evolved PDFs of the $K^{-}$ meson at $Q^2=100~\mathrm{GeV}^2$ are presented in Fig.~\ref{fig:kaon}. The uncertainty bands correspond to varying the initial scale within the range $Q_0$--$2Q_0$.

The NLO evolution exhibits the characteristic redistribution of the meson longitudinal momentum among the different partonic constituents. As the energy scale increases, repeated quark-to-gluon radiation followed by gluon splitting into quark-antiquark pairs progressively transfers momentum from the valence sector to the dynamically generated gluon and sea-quark distributions. Consequently, the valence PDFs become suppressed, particularly in the intermediate- and low-$x$ regions, whereas the gluon and sea-quark distributions increase rapidly at small $x$, consistent with perturbative QCD expectations.

Despite this continuous redistribution of momentum, the flavor hierarchy established at the model scale remains remarkably stable under QCD evolution. The strange quark continues to carry a larger longitudinal momentum fraction than the light antiquark over the entire evolution range because the heavier constituent is initially localized at larger $x$, and the DGLAP evolution modifies the distributions without altering this underlying mass hierarchy. This demonstrates that the constituent-mass dependence predicted by the LCQM is a robust non-perturbative feature that survives perturbative evolution to experimentally relevant energy scales.

The evolved PDFs satisfy the momentum sum rule at every energy scale,
\begin{align}
\int_{0}^{1} dx\, x
\Big[
f_{\mathrm{val}}(x,Q^{2})
+\bar{f}_{\mathrm{val}}(x,Q^{2})
\nonumber\\
+\, S(x,Q^{2})
+ g(x,Q^{2})
\Big]
=1.
\end{align}

The evolution of the average momentum fractions and the Mellin moments is also presented in Fig.~\ref{fig:kaon}. As the evolution proceeds, the reduction of the valence momentum fraction is exactly compensated by the growth of the gluon and sea-quark contributions, thereby preserving the total longitudinal momentum of the meson throughout the evolution.

The higher-order Mellin moments decrease monotonically with increasing $Q^2$, indicating that QCD evolution progressively broadens the longitudinal momentum distributions and reduces the probability of finding a constituent carrying a large fraction of the meson momentum. At the same time, the ordering of the Mellin moments among the individual constituents remains unchanged throughout the evolution, confirming that the mass-dependent momentum hierarchy established at the non-perturbative scale is preserved after NLO DGLAP evolution. The simultaneous evolution of the PDFs and their Mellin moments therefore provides a consistent picture of how perturbative QCD modifies the meson structure while retaining the essential non-perturbative characteristics of the underlying bound state.
At present, no direct experimental determination of the kaon PDFs is available. Therefore, lattice QCD calculations and phenomenological extractions provide the primary benchmarks for assessing the reliability of theoretical predictions \cite{Miller:2025wgr,Barry:2025wjx}. The evolved quark and antiquark PDFs are compared with the recent JAM analysis in Fig.~\ref{kaonjam} at the scale $Q^2=4~\mathrm{GeV}^2$, where the uncertainty band reflects the variation of the initial scale considered in this work. The evolved PDFs are found to be in good agreement with the JAM results in the large-$x$ region, where the distributions are dominated by valence quarks. Moderate deviations appear in the intermediate-$x$ region, while our PDFs are systematically larger than the recent phenomenological extraction of Ref.~\cite{Bourrely:2023yzi} and the lattice QCD results of Ref.~\cite{Miller:2025wgr}. These differences primarily originate from the choice of the initial non-perturbative scale and the corresponding valence distributions used as inputs to the QCD evolution. Nevertheless, the overall agreement demonstrates that the present LCQM provides a realistic description of the kaon valence structure after NLO DGLAP evolution.

The calculated Mellin moments of the kaon constituents up to $n=2$ are listed in Table~\ref{kaon_moments}. The valence constituents are found to carry larger average longitudinal momentum fractions than those predicted by the available theoretical models and lattice QCD calculations \cite{Lan:2019rba,Choi:2025bxk,Chen:2016sno,Watanabe:2018qju,Alexandrou:2021mmi}. For a more quantitative comparison, the normalized Mellin moment ratios, $\langle x^n\rangle/\langle x\rangle$, of the valence, gluon, and sea-quark distributions are presented in Fig.~\ref{kaonMellin} at $Q^2=4$, $16$, and $27~\mathrm{GeV}^2$. Unlike the individual Mellin moments, these normalized ratios are less sensitive to the overall normalization of the PDFs and therefore provide a more stringent test of the shape of the parton distributions. The calculated ratios are found to be in good agreement with the available theoretical models and lattice QCD simulations, indicating that the longitudinal momentum distributions predicted by the present framework are consistent with other non-perturbative approaches. The comparatively larger uncertainty bands originate predominantly from the variation of the initial evolution scale rather than from the perturbative evolution itself.

We further predict the kaon structure function at NLO for the kinematic region relevant to future EIC measurements. The structure function is calculated from the evolved quark, antiquark, and gluon distributions as
\begin{equation}
\begin{aligned}
F_{2}(x, Q^{2}) 
= \sum_{q} e_{q}^{2}\, x \Bigg\{ 
& f_q(x, Q^{2}) +  f_{\bar q}(x, Q^{2})  + \frac{\alpha_{s}(Q^{2})}{2\pi} 
 \\
& \times \Big[ 
C_{2q} [z] \otimes \big( f_q(x, Q^{2}) + f_{\bar q}(x, Q^{2}) \big)  \\
& \qquad\qquad + 2\, C_{2g} [z] \otimes g(x, Q^{2}) 
\Big] 
\Bigg\},
\end{aligned}
\end{equation}
where the coefficient functions in the $\overline{\mathrm{MS}}$ scheme are given by
\begin{equation}
\begin{aligned}
C_{2q}^{\overline{\mathrm{MS}}}(z) =\;
& C_F \Bigg[
2 \left( \frac{\ln(1 - z)}{1 - z} \right)_{+}
- \frac{3}{2} \left( \frac{1}{1 - z} \right)_{+} \\
&\quad - (1 + z)\ln(1 - z)
- \frac{1 + z^2}{1 - z} \ln z \\
&\quad + 3 + 2z
- \left( \frac{\pi^2}{3} + \frac{9}{2} \right)\delta(1 - z)
\Bigg],
\end{aligned}
\end{equation}
and
\begin{equation}
\begin{aligned}
C_{2g}^{\overline{\mathrm{MS}}}(z) =\;
& T_R \Big[
\big( (1 - z)^2 + z^2 \big)
\ln\!\left(\frac{1 - z}{z}\right) \\
&\quad - 8z^2 + 8z - 1
\Big].
\end{aligned}
\end{equation}
Here, $C_F=4/3$ and $T_R=1/2$ are the color factors. In Fig.~\ref{structureF2}, we have plotted the kaon structure function as a function of $x$ at the energy scales $Q^2=10$, $100$, and $1000~\mathrm{GeV}^2$. The structure function is found to increase in the small-$x$ region ($x\lesssim0.2$) with increasing $Q^2$, while it decreases in the large-$x$ region. This behavior arises from the transfer of longitudinal momentum from the valence quarks to the dynamically generated gluon and sea-quark distributions during QCD evolution. The predicted kaon structure function will be useful for future EIC measurements, where the kaon structure will be explored through the Sullivan process in tagged deep-inelastic scattering (TDIS) \cite{JLabC12-15-006A}.
\par We have also predicted the differential cross section for the kaon-induced fixed-target Drell--Yan process ($K^{\pm}+A\rightarrow \ell^{+}\ell^{-}+X$) relevant to the upcoming COMPASS++/AMBER experiment at CERN. The COMPASS++/AMBER experiment will measure the Drell--Yan cross section using a kaon beam incident on a carbon target, which contains equal numbers of protons and neutrons, to investigate the partonic structure of the kaon \cite{Adams:2018pwt}. Besides carbon (C), we also present predictions for aluminum (Al) and tungsten (W) nuclear targets using the corresponding CTEQ15 nuclear PDFs \cite{Kovarik:2015cma}. The differential cross section for the kaon-induced Drell--Yan process at NLO is given by \cite{Lan:2019rba,Anastasiou:2003yy,P:2026crg}
\begin{equation}
\begin{aligned}
\frac{m^3 d^2\sigma}{dm\,dY}
&=
\frac{8\pi\alpha_s^2(Q^{2})}{9}
\frac{m^2}{s}
\sum_{ij}
\int dx_1\,dx_2
\\
&\times
C_{ij}(x_1,x_2,s,m,Q^2)
f_{i/K}(x_1,Q^2)
f_{j/A}(x_2,Q^2),
\end{aligned}
\end{equation}
where $C_{ij}$ represents the hard-scattering kernels, which are expanded perturbatively as
\begin{equation}
C_{ij}(x_1,x_2,s,m,Q^2)
=
\sum_{n=0}^{\infty}
\left(\frac{\alpha_s}{2\pi}\right)^n
C_{ij}^{(n)}(x_1,x_2,s,m,Q^2).
\end{equation}
The NLO coefficient functions $C_{ij}^{(n)}$ have been taken from Ref.~\cite{Anastasiou:2003yy}. The summation includes the $q\bar q$, $qg$, and $\bar q g$ subprocesses. Here, $f_{i/K}(x_1,Q^2=m^2)$ denotes the evolved kaon PDF, while $f_{j/A}(x_2,Q^2)$ represents the nuclear PDF of the target. By integrating over the rapidity $Y$, we obtain the differential cross section, which is shown in Fig.~\ref{dykaon}. We have considered both $K^{+}(u\bar s)$ and $K^{-}(s\bar u)$ beams with the same beam energy of $100~\mathrm{GeV}$. The $K^{-}$ beam is found to produce a larger cross section than the $K^{+}$ beam, which is consistent with the trend observed in kaon-induced $J/\psi$ production \cite{Corden:1980rb}. The calculated cross sections also exhibit behavior similar to the recent BLFQ predictions \cite{Lan:2025fia}, while the choice of nuclear target is found to have only a modest effect on the total cross section. By integrating the differential cross sections over the considered di-lepton invariant-mass range, we obtain the integrated cross-section ratio
\[
\frac{\sigma(K^+)}{\sigma(K^-)}
=
0.28\pm0.012,
\]
which is in good agreement with the ratio measured in kaon-induced $J/\psi$ production,
$\sigma(K^+)/\sigma(K^-)=0.29\pm0.07$ \cite{Corden:1980rb}. Although these two processes involve different production mechanisms, the similar ratio indicates that the suppression of the $K^{+}$ cross section relative to the $K^{-}$ cross section is primarily governed by the different valence-quark compositions of the two kaon beams. Future COMPASS++/AMBER measurements will provide a direct test of this prediction.
\begin{table*}[t]
\centering
\caption{Valence quark momentum asymmetry 
$\langle x \rangle_q - \langle x \rangle_{\bar q}$ 
at $Q^2 = 10^{4}\,\mathrm{GeV}^2$ for different meson.}
\begin{ruledtabular}
\begin{tabular}{cccc}
Meson & $\langle x \rangle_{\text{quark}}$ 
& $\langle x \rangle_{\text{antiquark}}$ 
& $\langle x \rangle_q - \langle x \rangle_{\bar q}$ \\
\hline

$K^{-}$  
& $0.276^{+0.029}_{-0.061}$ 
& $0.228^{+0.024}_{-0.048}$ 
& $0.048^{+0.006}_{-0.012}$ \\

$D^{+}$ 
& $0.484^{+0.023}_{-0.036}$ 
& $0.197^{+0.009}_{-0.013}$ 
& $0.287^{+0.014}_{-0.023}$ \\

$D_{s}$  
& $0.459^{+0.022}_{-0.034}$ 
& $0.222^{+0.010}_{-0.015}$ 
& $0.237^{+0.012}_{-0.019}$ \\

$\eta_{c}$  
& $0.340^{+0.016}_{-0.025}$ 
& $0.341^{+0.016}_{-0.025}$ 
& $0.0$ \\

$B_{s}^{0}$  
& $0.134^{+0.004}_{-0.006}$ 
& $0.660^{+0.022}_{-0.033}$ 
& $-0.526^{+0.027}_{-0.018}$ \\

$B_{c}$  
& $0.232^{+0.007}_{-0.010}$ 
& $0.562^{+0.019}_{-0.028}$ 
& $-0.331^{+0.017}_{-0.012}$ \\

$\eta_{b}$ 
& $0.397^{+0.013}_{-0.019}$ 
& $0.397^{+0.013}_{-0.019}$ 
& $0.0$ \\

$B^{+}$ 
& $0.117^{+0.003}_{-0.005}$ 
& $0.678^{+0.023}_{-0.034}$ 
& $-0.559^{+0.029}_{-0.019}$ \\
\end{tabular}
\end{ruledtabular}
\label{asymmetryMellin}
\end{table*}
\par Except for the kaon, direct experimental information on the internal structure of the heavier mesons considered in this work is currently unavailable because of their extremely short lifetimes. Nevertheless, understanding their PDFs is important for describing heavy-ion collisions, quarkonium production, and various exclusive and inclusive hard-scattering processes, where non-perturbative QCD effects play an essential role. In this work, we have evolved the constituent PDFs over the range $x\ge10^{-5}$ to energy scales relevant for the proposed Electron-Ion Collider in China (EicC) \cite{Chen:2018wyz}, the electron--Relativistic Heavy Ion Collider (eRHIC) \cite{Aschenauer:2014cki}, and the Large Hadron Electron Collider (LHeC) \cite{LHeCStudyGroup:2012zhm}.

The detailed evolution of the heavy-meson PDFs shown in Figs.~\ref{dsplus}, \ref{etac}, \ref{etab}, and \ref{bplus} has been presented in the Appendix to avoid repetition, since all heavy mesons exhibit the same qualitative features under NLO DGLAP evolution. The principal differences arise only from the constituent masses and the corresponding initial evolution scales.

For the heavy--light systems ($D$, $D_s$, $B$, $B_s$, and $B_c$), the heavier constituent consistently carries a larger longitudinal momentum fraction than the lighter constituent over the entire evolution range. Compared with the kaon, these mesons generate smaller gluon and sea-quark distributions because of their larger initial scales and the correspondingly shorter DGLAP evolution length. As a result, the valence sector remains dominant even at high momentum scales.

For the quarkonium systems ($\eta_c$ and $\eta_b$), the quark and antiquark carry identical momentum fractions owing to their equal constituent masses. The dynamically generated heavy-flavor sea enhances the distributions only in the very small-$x$ region, while the valence contribution remains dominant throughout the evolution. The bottomonium distributions are found to be narrower than the charmonium distributions, reflecting the larger bottom-quark mass.

The bottom-meson family exhibits the strongest constituent-mass asymmetry. In particular, the bottom antiquark carries most of the longitudinal momentum in the $B^+$ and $B_s^0$ mesons, whereas this dominance is reduced in the $B_c$ meson because the accompanying charm quark is substantially heavier than the light up or strange quarks. These features are consistent with recent BLFQ calculations for heavy mesons \cite{Lan:2019img}.

\par For an overall comparison, the evolved valence, gluon, and sea-quark distributions of all mesons at $Q^2=10^4~\mathrm{GeV}^2$ are presented in Fig.~\ref{allmesons}. The qualitative mass dependence established at the model scale is preserved after evolution, with heavier constituents remaining concentrated at larger values of $x$ than lighter constituents. The kaon exhibits the largest gluon and sea-quark distributions among all mesons considered, whereas these distributions become progressively smaller with increasing constituent masses. Both the gluon and sea-quark distributions rapidly decrease for $x\gtrsim0.6$.

The corresponding average momentum fractions are shown in Fig.~\ref{allmmvalue}. The heavy mesons retain a significantly larger fraction of their momentum in the valence sector than the kaon. In particular, the valence constituents of the bottomonium and $B$-meson systems carry approximately $75\%$--$83\%$ of the total longitudinal momentum, whereas the corresponding fraction for the kaon is considerably smaller. Consequently, the gluon and sea-quark momentum fractions remain below $10\%$ for all heavy mesons considered in this work.

The Mellin moments of the evolved quark and antiquark PDFs are summarized in Fig.~\ref{fig:Mellin_momentsall} and in Table~\ref{asymmetryMellin}. The first Mellin moments preserve the same ordering observed at the initial scale [Eq.~(\ref{ruleee})], whereas the higher-order moments evolve differently owing to QCD radiation. The quark--antiquark momentum asymmetry,
$\langle x\rangle_q-\langle x\rangle_{\bar q}$,
is presented in Figs.~\ref{fixmelli} and \ref{differenceMellin}, with the corresponding numerical values at $Q^2=10^4$ GeV$^2$ listed in Table~\ref{asymmetryMellin}. Although the magnitude of the asymmetry decreases gradually with increasing $Q^2$, its dependence on the constituent-mass difference remains unchanged throughout the evolution. The largest asymmetry is obtained for the $B^+$ meson because of the large mass difference between the bottom and up quarks, whereas the kaon exhibits only a weak scale dependence. The magnitude of the quark--antiquark momentum asymmetry follows the hierarchy
\begin{equation}
| \langle x \rangle_q-\langle x \rangle_{\bar q}|
:
B^+
>
B_s^0
>
B_c
>
D^+
>
D_s
>
K^-.
\end{equation}
We have also compared the bottom- and charm-quark PDFs in Figs.~\ref{bquark} and \ref{cquark}, respectively. The bottom PDFs exhibit distinct distributions among different mesons, with the distributions becoming increasingly concentrated at large $x$ owing to the heavy bottom-quark mass. As the mass of the accompanying constituent decreases, the peak gradually shifts towards lower values of $x$. A similar trend is observed for the charm-quark distributions. The top sea-quark PDFs, shown in Fig.~\ref{top}, are evaluated at $Q^2=10^5~\mathrm{GeV}^2>m_t^2$, where they are generated perturbatively through gluon splitting, $g\rightarrow t\bar{t}$. Among all the mesons considered, the kaon exhibits the largest top sea-quark distribution because its larger gluon density after DGLAP evolution enhances heavy-flavor production.
\par Overall, the PDFs of different mesons exhibit distinct structural features that reflect their flavor compositions and underlying non-perturbative dynamics. To examine the dependence of our results on the choice of the initial evolution scale, we also performed the QCD evolution using a common initial scale for all mesons. Both the qualitative behavior and the numerical predictions remain largely unchanged, demonstrating that the observed differences among the meson PDFs originate primarily from their intrinsic non-perturbative structures rather than from the choice of the initial evolution scale.

\section{Conclusion}
\label{conclus}
In this work, we have investigated the parton distribution functions (PDFs) of pseudoscalar mesons at the model scale and after QCD evolution to higher energy scales. The quark and antiquark PDFs at the initial scale were calculated within the light-cone quark model (LCQM) using the Brodsky--Huang--Lepage momentum-space wave functions. By solving the quark--quark correlation functions, we obtained explicit expressions for the valence PDFs. The PDFs were subsequently evolved to higher momentum scales using the next-to-leading-order (NLO) DGLAP evolution equations implemented through the HOPPET toolkit. The calculated PDFs satisfy the required quark-number and momentum sum rules.

A systematic study of the kaon, $D$-mesons, $B$-mesons, and heavy quarkonia demonstrates that the constituent mass plays the dominant role in determining the longitudinal momentum distributions. Heavy constituents consistently carry a larger fraction of the meson momentum than light constituents, and this hierarchy is preserved under NLO DGLAP evolution. Although QCD evolution dynamically generates gluon and sea-quark distributions, the qualitative mass dependence established at the model scale remains unchanged over the entire evolution range.

The PDFs corresponding to heavier constituent quarks are found to peak in the large-$x$ region, whereas those of lighter constituents are broader and shifted towards smaller values of $x$. Mesons composed of equal-mass constituents exhibit symmetric PDFs about $x=0.5$, while asymmetric systems display a pronounced momentum asymmetry between the quark and antiquark. Heavy mesons generate considerably smaller gluon and sea-quark distributions than the kaon because of their larger initial scales and shorter evolution lengths.

The evolved Mellin moments further demonstrate that the valence sector remains dominant in heavy mesons even at high energy scales, whereas the gluon and sea-quark momentum fractions become increasingly important for the kaon. The quark--antiquark momentum asymmetry is found to be strongly correlated with the constituent-mass difference and decreases only gradually with increasing $Q^2$.

For the kaon, we have presented predictions for the NLO structure function relevant to future Electron--Ion Collider measurements through tagged deep-inelastic scattering, together with NLO fixed-target Drell--Yan cross sections for $K^{+}$- and $K^{-}$induced reactions on carbon, aluminum, and tungsten targets relevant to the COMPASS++/AMBER experiment at CERN. The predicted integrated cross-section ratio, $\sigma(K^{+})/\sigma(K^{-})=0.28\pm0.012$, provides a quantitative benchmark for future experimental measurements. The calculated Mellin moments are found to be consistent with existing theoretical models and lattice QCD calculations.

Finally, although direct experimental information on heavy-meson PDFs remains limited, the present predictions provide a comprehensive theoretical benchmark for future phenomenological studies, lattice QCD calculations, and possible experimental investigations of heavy-meson structure.

\section*{Acknowledgment}
The author sincerely acknowledges valuable discussions and insightful suggestions from Craig D. Roberts, Wen-Chen Chang, Chia-Yu Hsieh, Marco Meyer-Conde, Narinder Kumar, Ashutosh Dwibedi, and Shubham Sharma. The author especially thanks Ashutosh Dwibedi and Shubham Sharma for carefully reading the manuscript and providing useful comments and suggestions.

\section{Appendix}
\begin{figure*}[t]
\centering
\begin{subfigure}{0.32\textwidth}
\centering
\includegraphics[width=\linewidth]{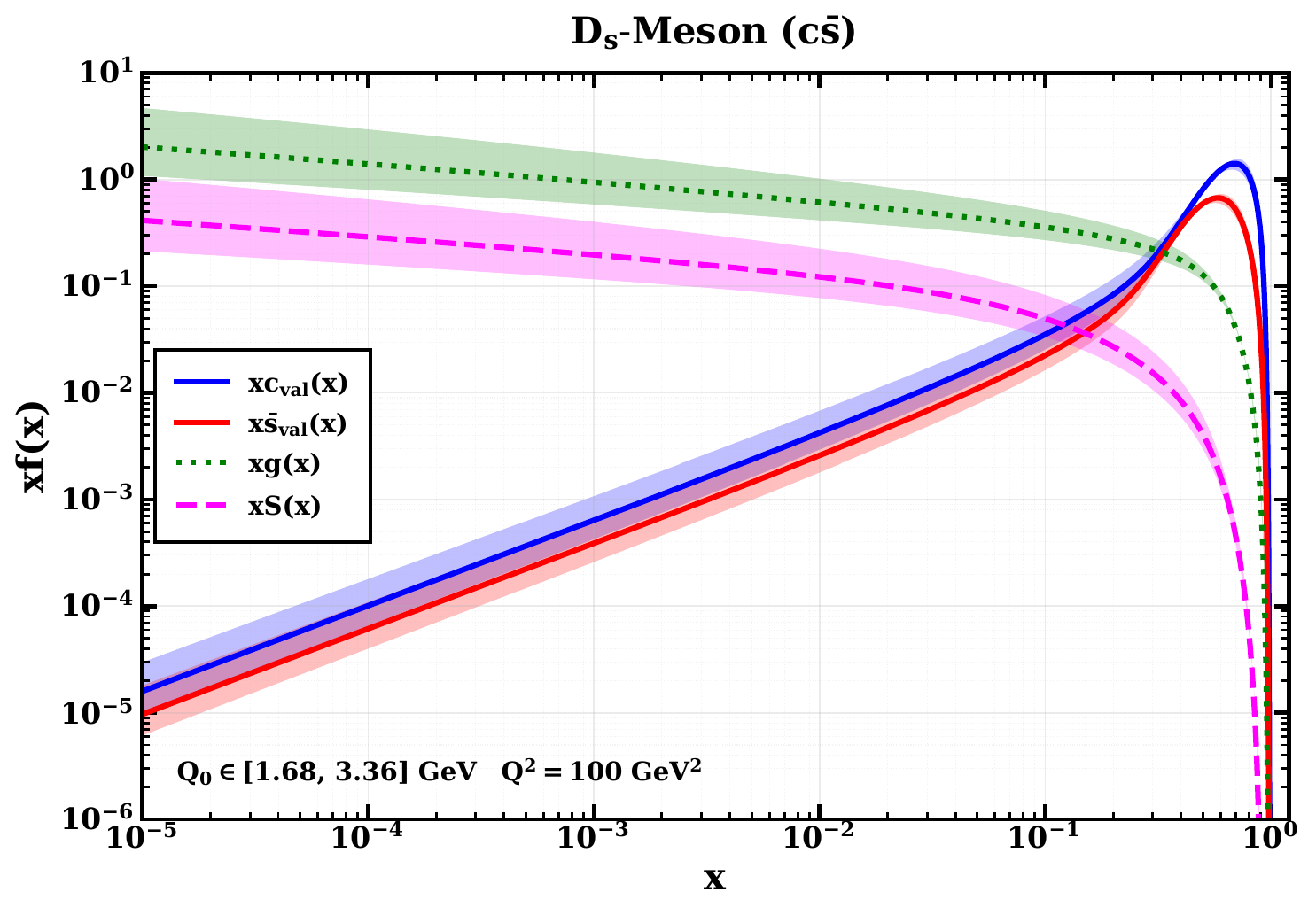}
\caption{The valence quark $xc_{val}(x)$, antiquark $x \bar s_{val}(x)$, gluon $x g(x)$, and sea-quark $xS(x)$ with respect to $x$.}
\end{subfigure}
\hfill
\begin{subfigure}{0.32\textwidth}
\centering
\includegraphics[width=\linewidth]{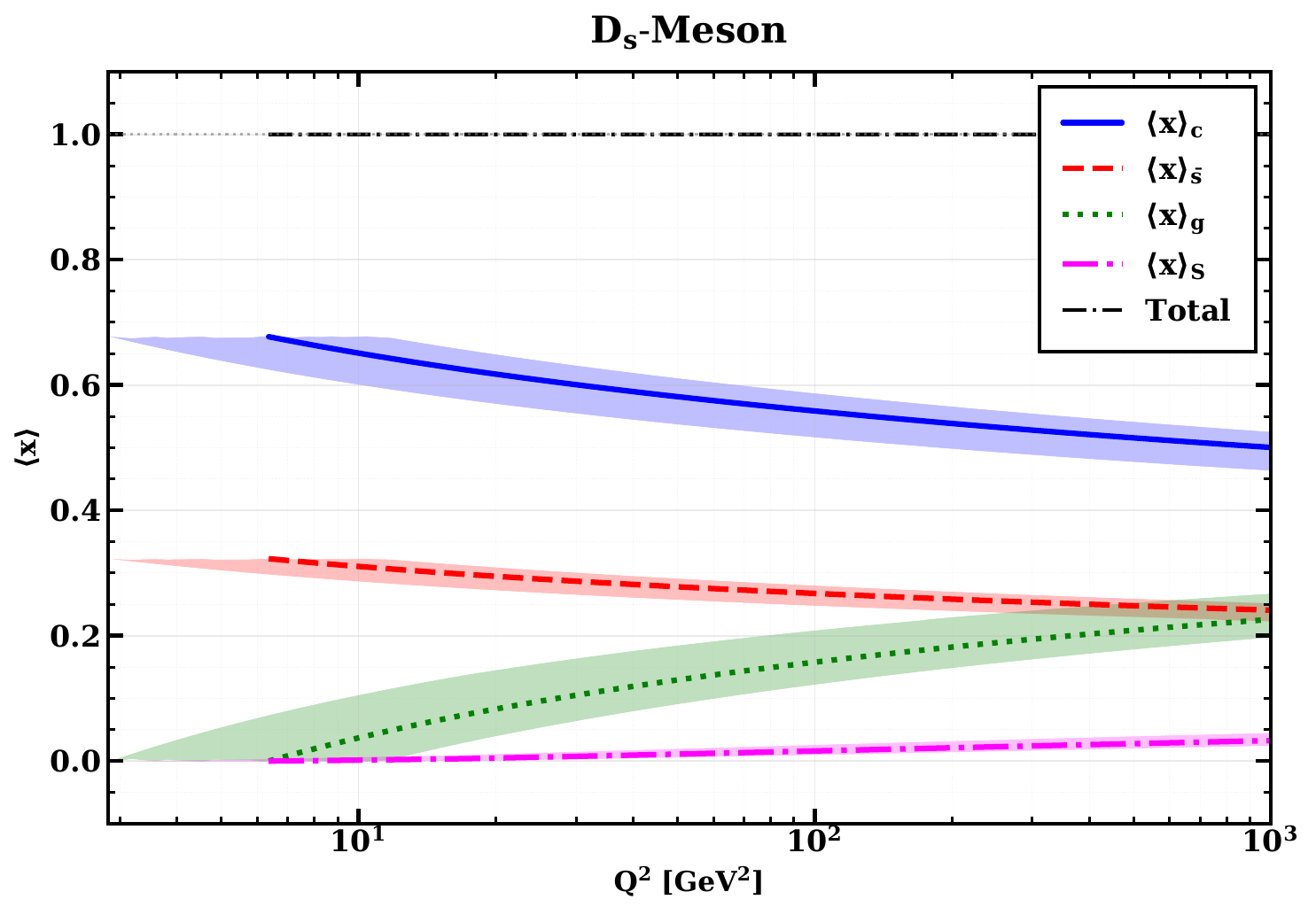}
\caption{Average momentum fractions carried by the quark $\langle x \rangle_c$, antiquark $\langle x \rangle_{\bar s}$, gluon $\langle x \rangle_g$ and sea-quarks $\langle x \rangle_S$ with respect to $Q^2$ GeV$^2$.}
\end{subfigure}
\hfill
\begin{subfigure}{0.32\textwidth}
\centering
\includegraphics[width=\linewidth]{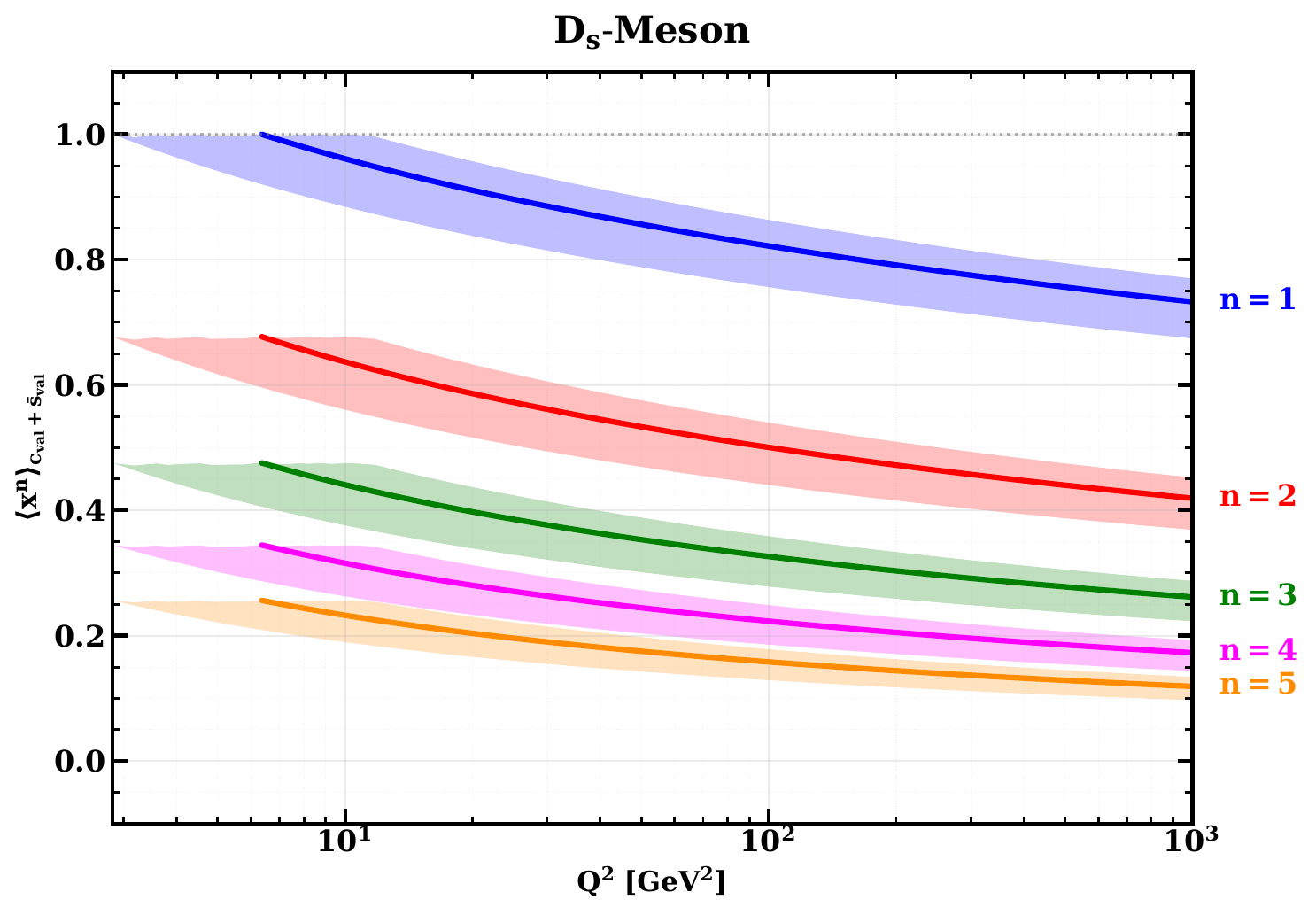}
\caption{The $n^{th}$ Mellin moment of the sum valence quark and antiquark $\langle x \rangle_{c_{val}+\bar s_{val}}$ with respect to $Q^2$ GeV$^2$ up to n=5.}
\end{subfigure}
\caption{The $D_s(c \bar s)$-meson PDFs have been evolved to $Q^2=100$ GeV$^2$ through NLO DGLAP evolutions by taking the initial scale $Q_0\in [1.68,3.36]$ GeV. The central line corresponds to the initial scale $Q_0= 2.52$ GeV.}
\label{dsplus}
\end{figure*}

\begin{figure*}[t]
\centering
\begin{subfigure}{0.32\textwidth}
\centering
\includegraphics[width=\linewidth]{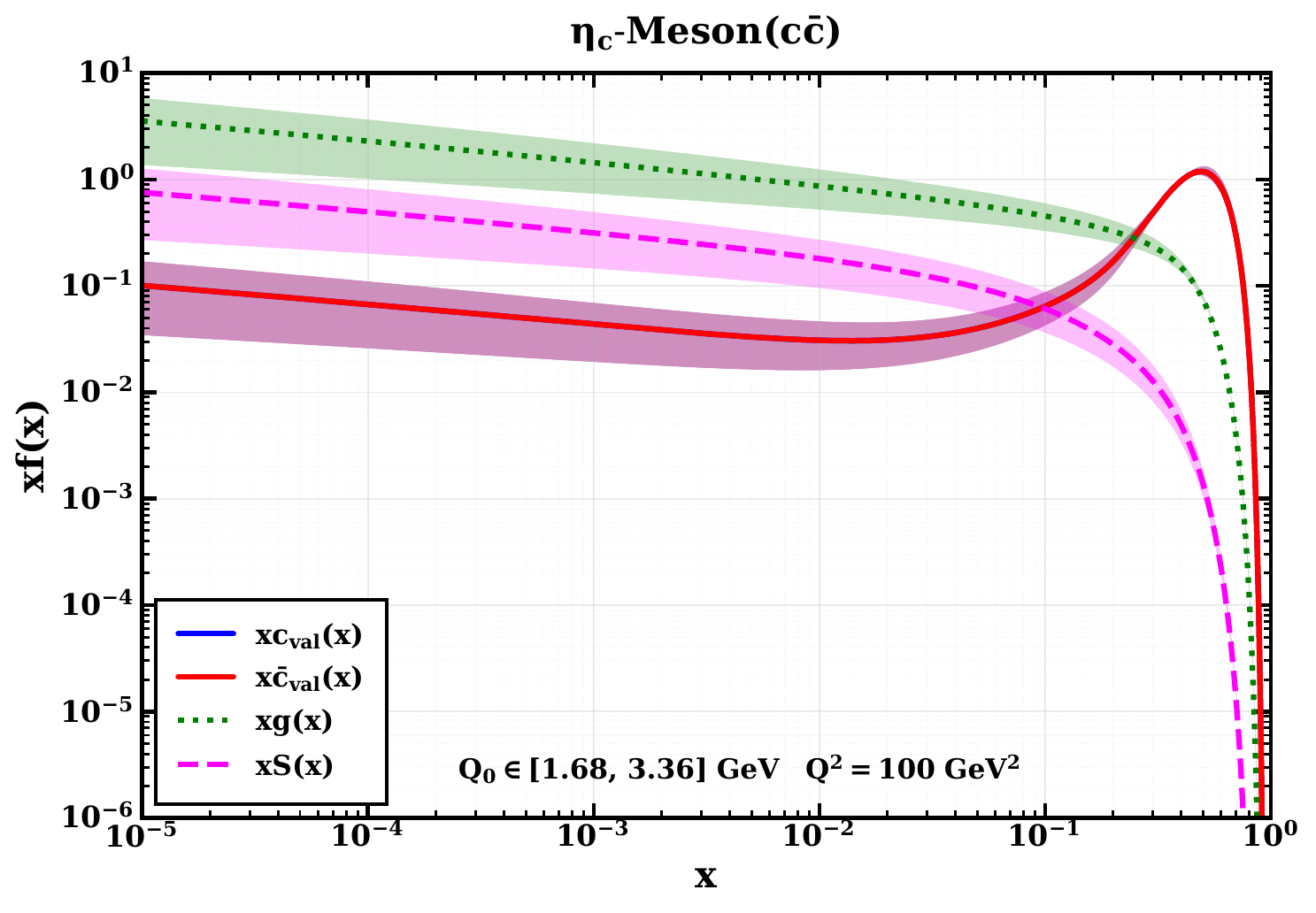}
\caption{The valence quark $xc_{val}(x)$, antiquark $x \bar c_{val}(x)$, gluon $x g(x)$, and sea-quark $xS(x)$ with respect to $x$.}
\end{subfigure}
\hfill
\begin{subfigure}{0.32\textwidth}
\centering
\includegraphics[width=\linewidth]{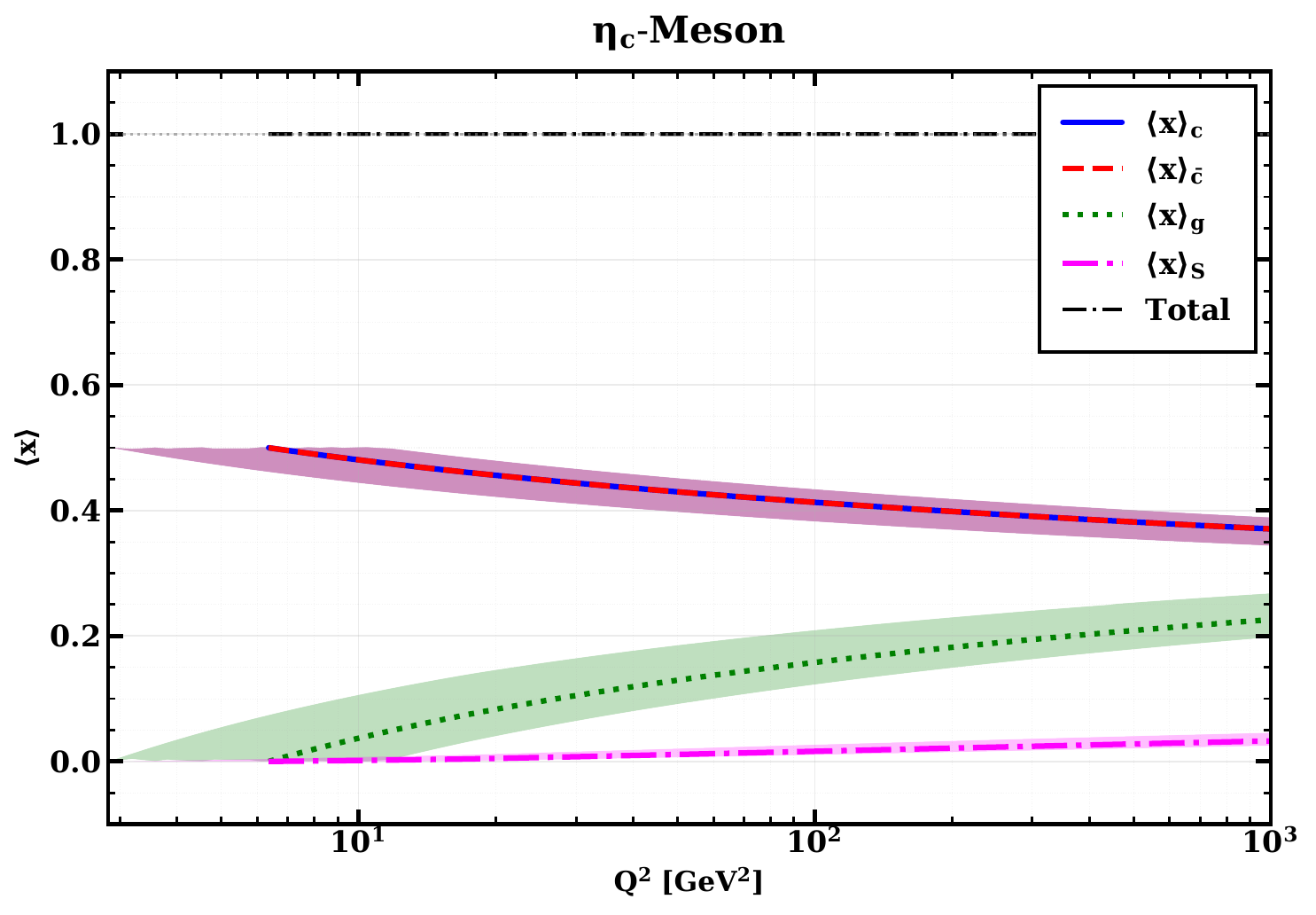}
\caption{Average momentum fractions carried by the quark $\langle x \rangle_c$, antiquark $\langle x \rangle_{\bar c}$, gluon $\langle x \rangle_g$ and sea-quarks $\langle x \rangle_S$ with respect to $Q^2$ GeV$^2$.}
\end{subfigure}
\hfill
\begin{subfigure}{0.32\textwidth}
\centering
\includegraphics[width=\linewidth]{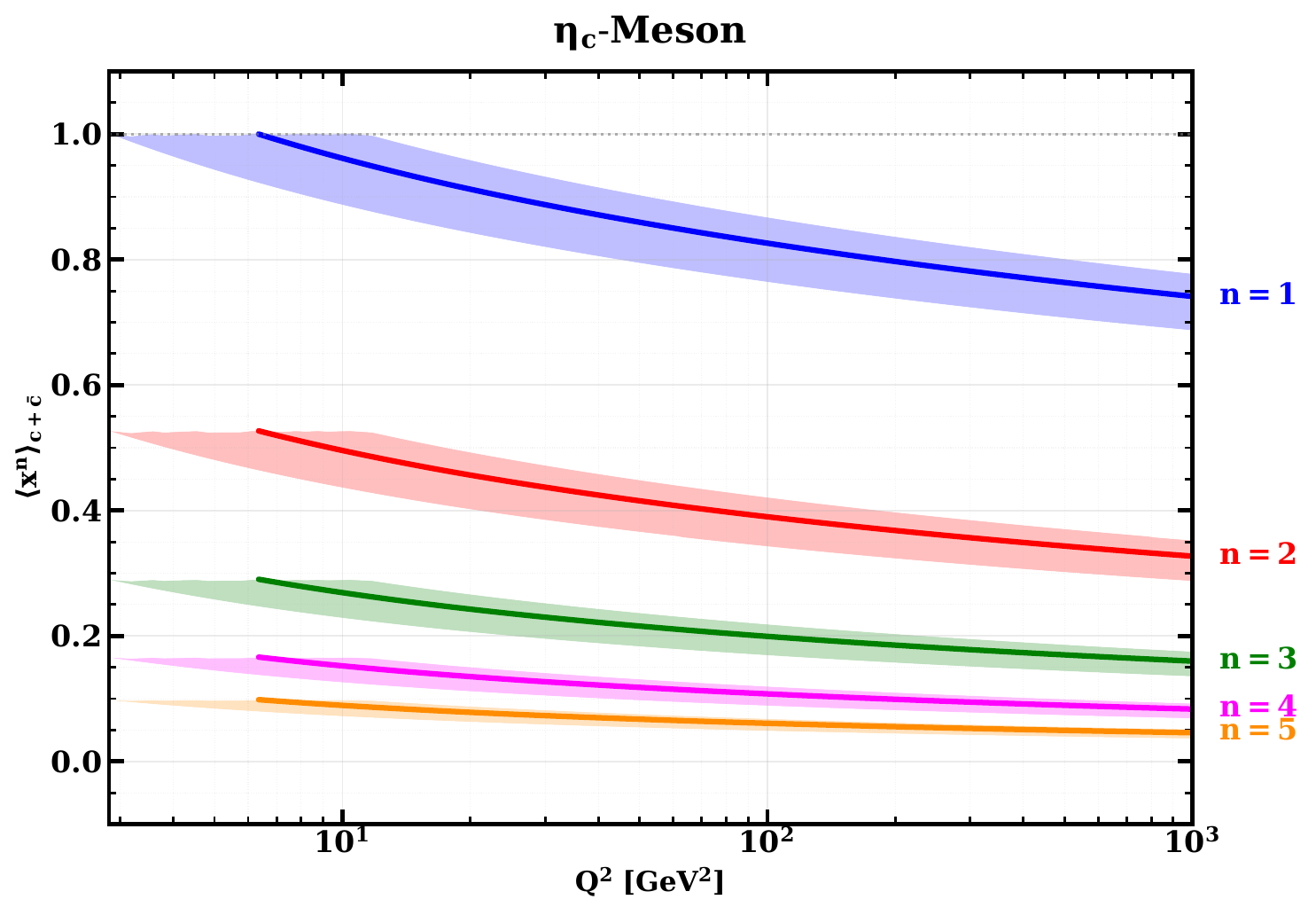}
\caption{The $n^{th}$ Mellin moment of the sum valence quark and antiquark $\langle x \rangle_{c_{val}+\bar c_{val}}$ with respect to $Q^2$ GeV$^2$ up to n=5.}
\end{subfigure}
\caption{(Color online) The $\eta_c(c \bar c)$-meson PDFs have been evolved to $Q^2=100$ GeV$^2$ through NLO DGLAP evolutions by taking the initial scale $Q_0\in [1.68,3.36]$ GeV. The central line corresponds to the initial scale $Q_0= 2.52$ GeV.}
\label{etac}
\end{figure*}
\begin{figure*}[t]
\centering
\begin{subfigure}{0.32\textwidth}
\centering
\includegraphics[width=\linewidth]{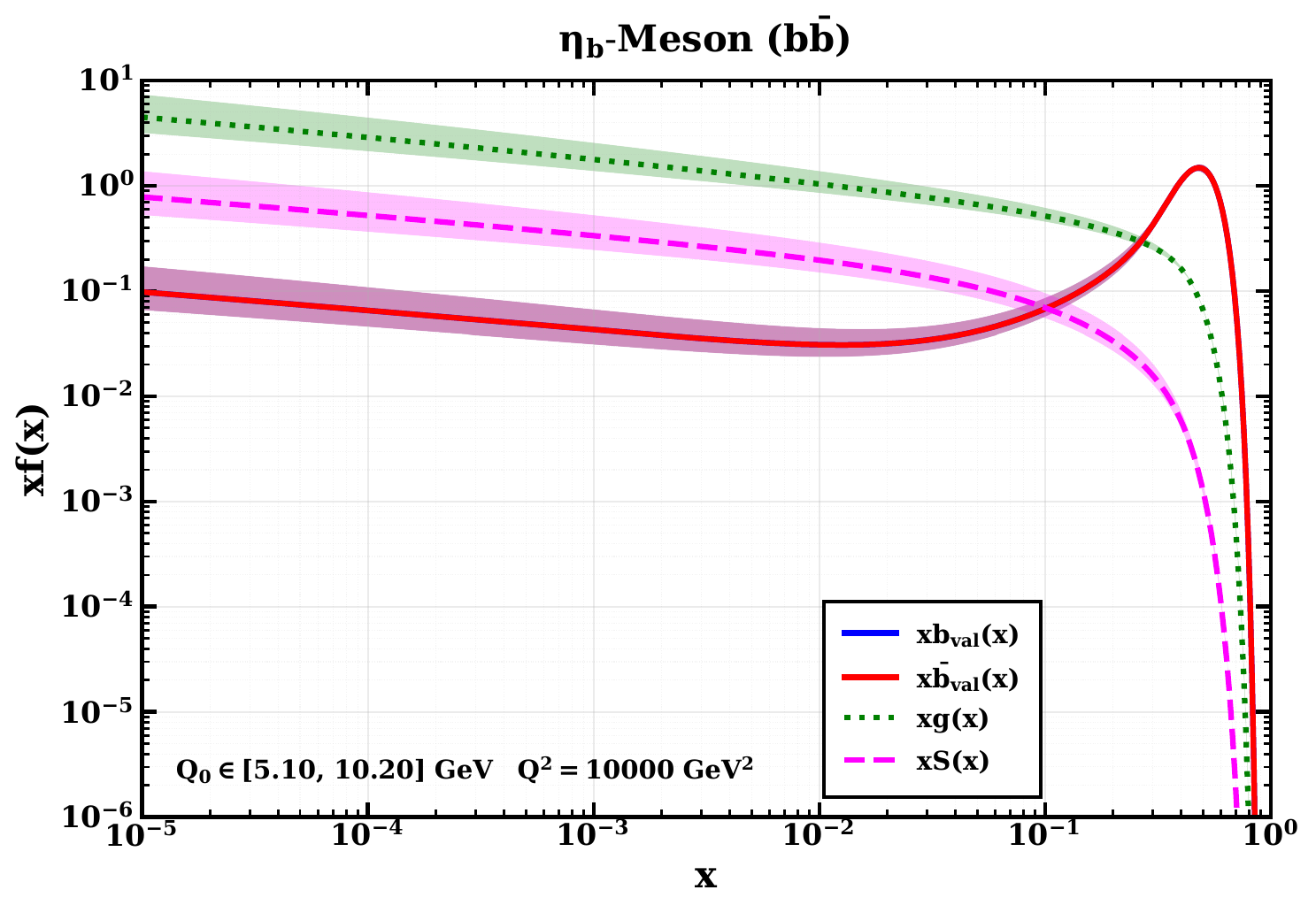}
\caption{The valence quark $xb_{val}(x)$, antiquark $x \bar b_{val}(x)$, gluon $x g(x)$, and sea-quark $xS(x)$ with respect to $x$.}
\end{subfigure}
\hfill
\begin{subfigure}{0.32\textwidth}
\centering
\includegraphics[width=\linewidth]{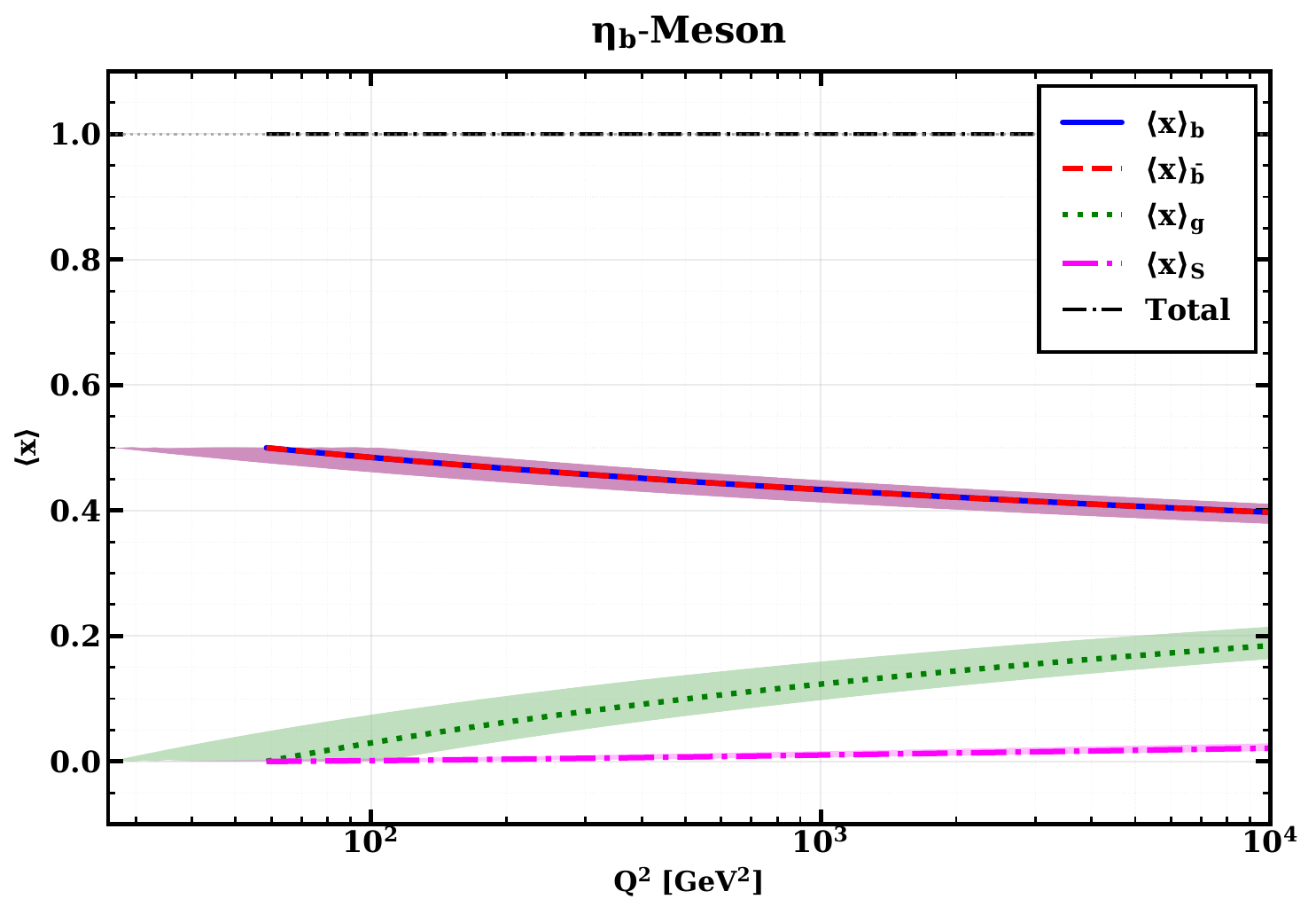}
\caption{Average momentum fractions carried by the quark $\langle x \rangle_b$, antiquark $\langle x \rangle_{\bar b}$, gluon $\langle x \rangle_g$ and sea-quarks $\langle x \rangle_S$ with respect to $Q^2$ GeV$^2$.}
\end{subfigure}
\hfill
\begin{subfigure}{0.32\textwidth}
\centering
\includegraphics[width=\linewidth]{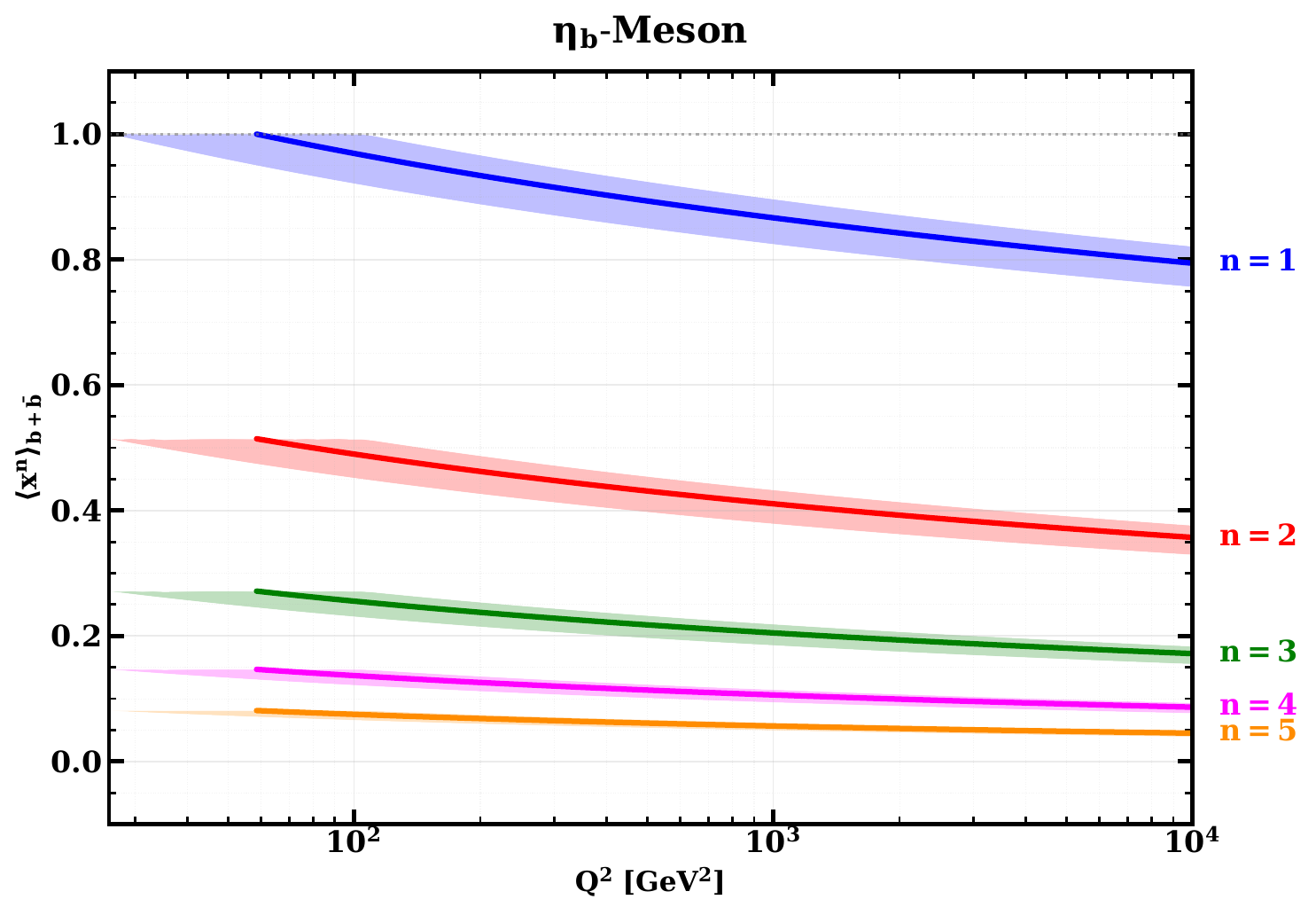}
\caption{The $n^{th}$ Mellin moment of the sum valence quark and antiquark $\langle x \rangle_{b_{val}+\bar b_{val}}$ with respect to $Q^2$ GeV$^2$ up to n = 5.}
\end{subfigure}
\caption{(Color online) The $\eta_b(b \bar b)$-meson PDFs have been evolved to $Q^2=1000$ GeV$^2$ through NLO DGLAP evolutions by taking the initial scale $Q_0\in [5.10,10.20]$ GeV. The central line corresponds to the initial scale $Q_0= 7.65$ GeV.}
\label{etab}
\end{figure*}
\begin{figure*}[t]
\centering
\begin{subfigure}{0.32\textwidth}
\centering
\includegraphics[width=\linewidth]{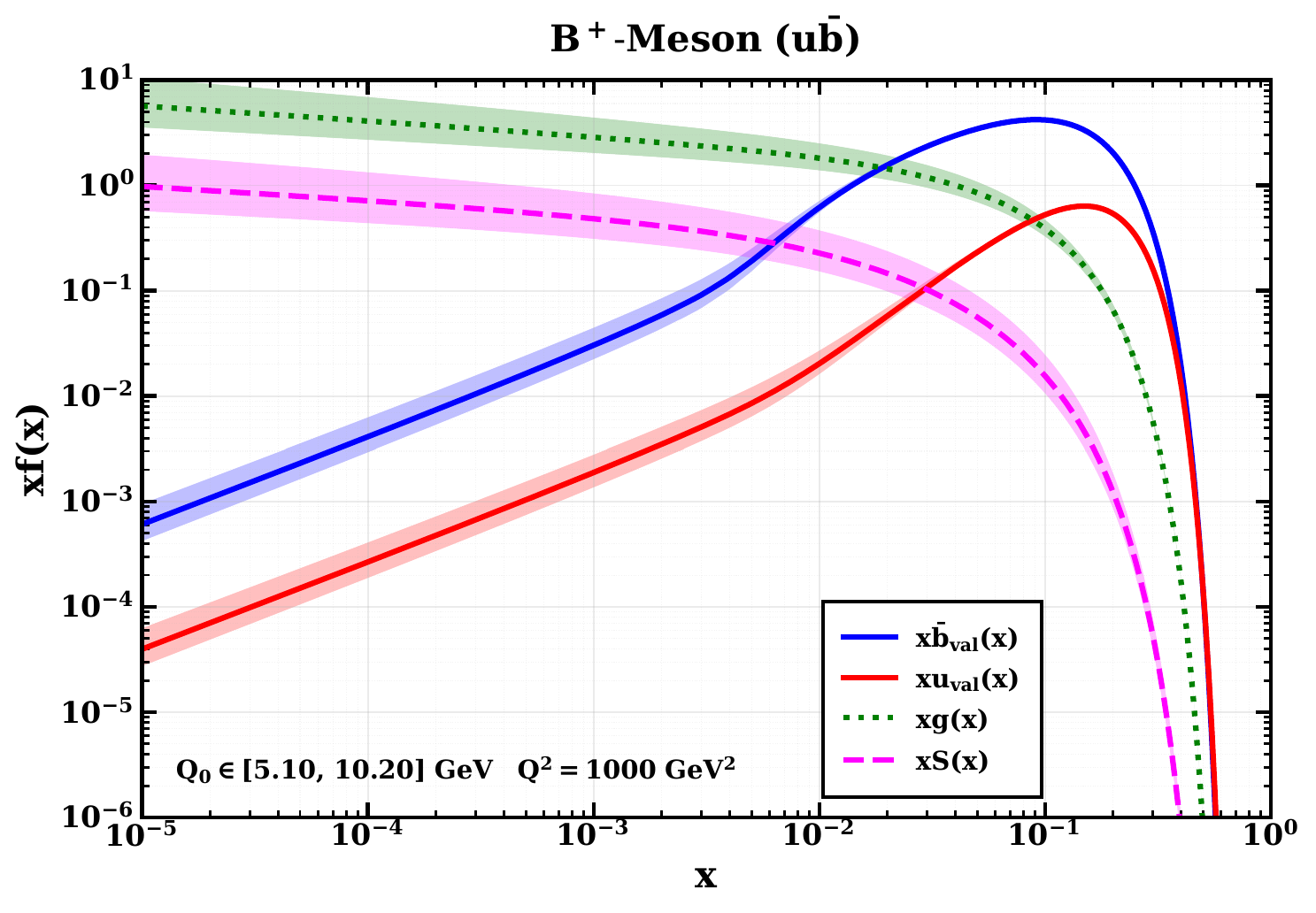}
\caption{The valence quark $xu_{val}(x)$, antiquark $x \bar b_{val}(x)$, gluon $x g(x)$, and sea-quark $xS(x)$ with respect to $x$.}
\end{subfigure}
\hfill
\begin{subfigure}{0.32\textwidth}
\centering
\includegraphics[width=\linewidth]{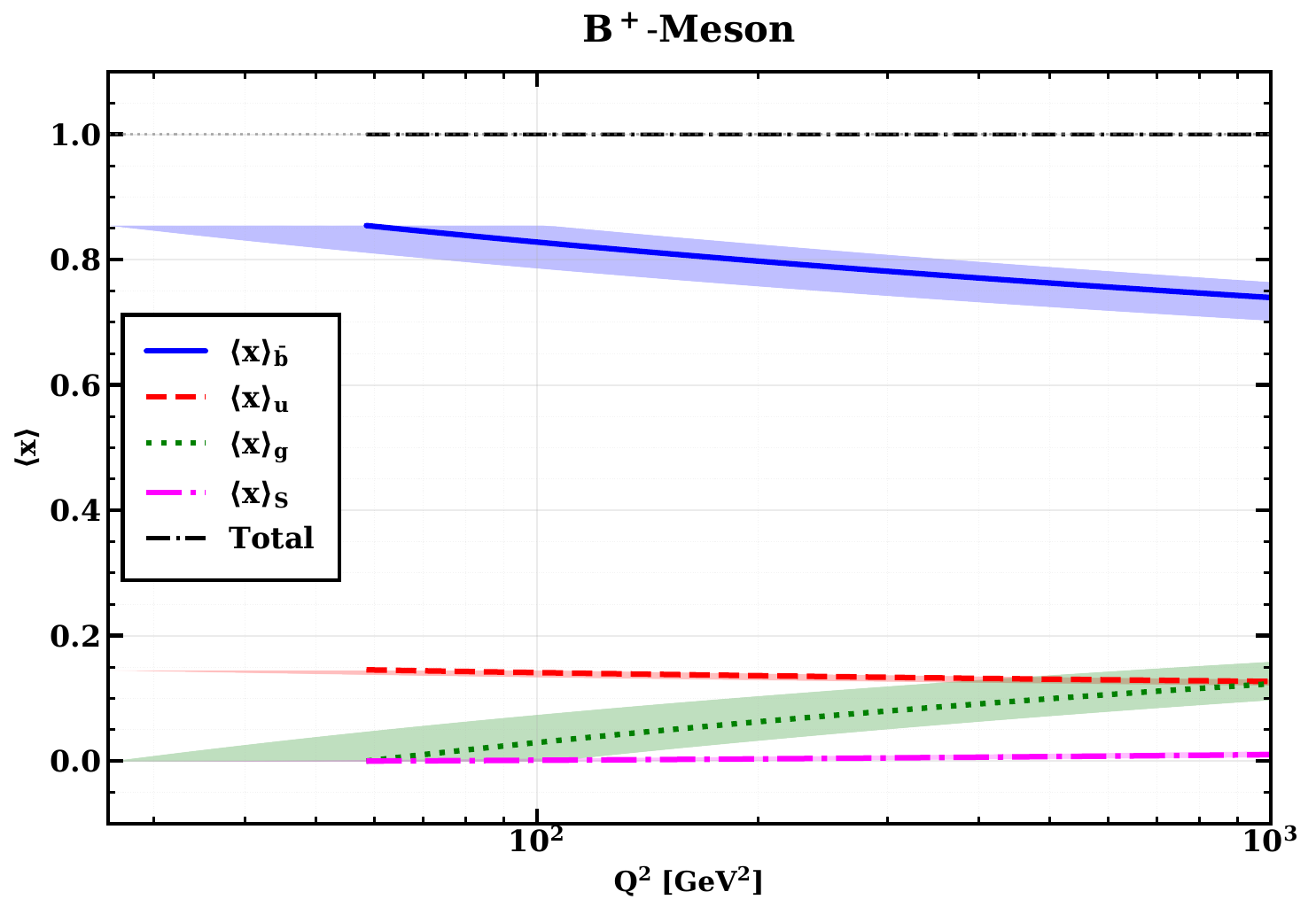}
\caption{Average momentum fractions carried by the quark $\langle x \rangle_u$, antiquark $\langle x \rangle_{\bar b}$, gluon $\langle x \rangle_g$ and sea-quarks $\langle x \rangle_S$ with respect to $Q^2$ GeV$^2$.}
\end{subfigure}
\hfill
\begin{subfigure}{0.32\textwidth}
\centering
\includegraphics[width=\linewidth]{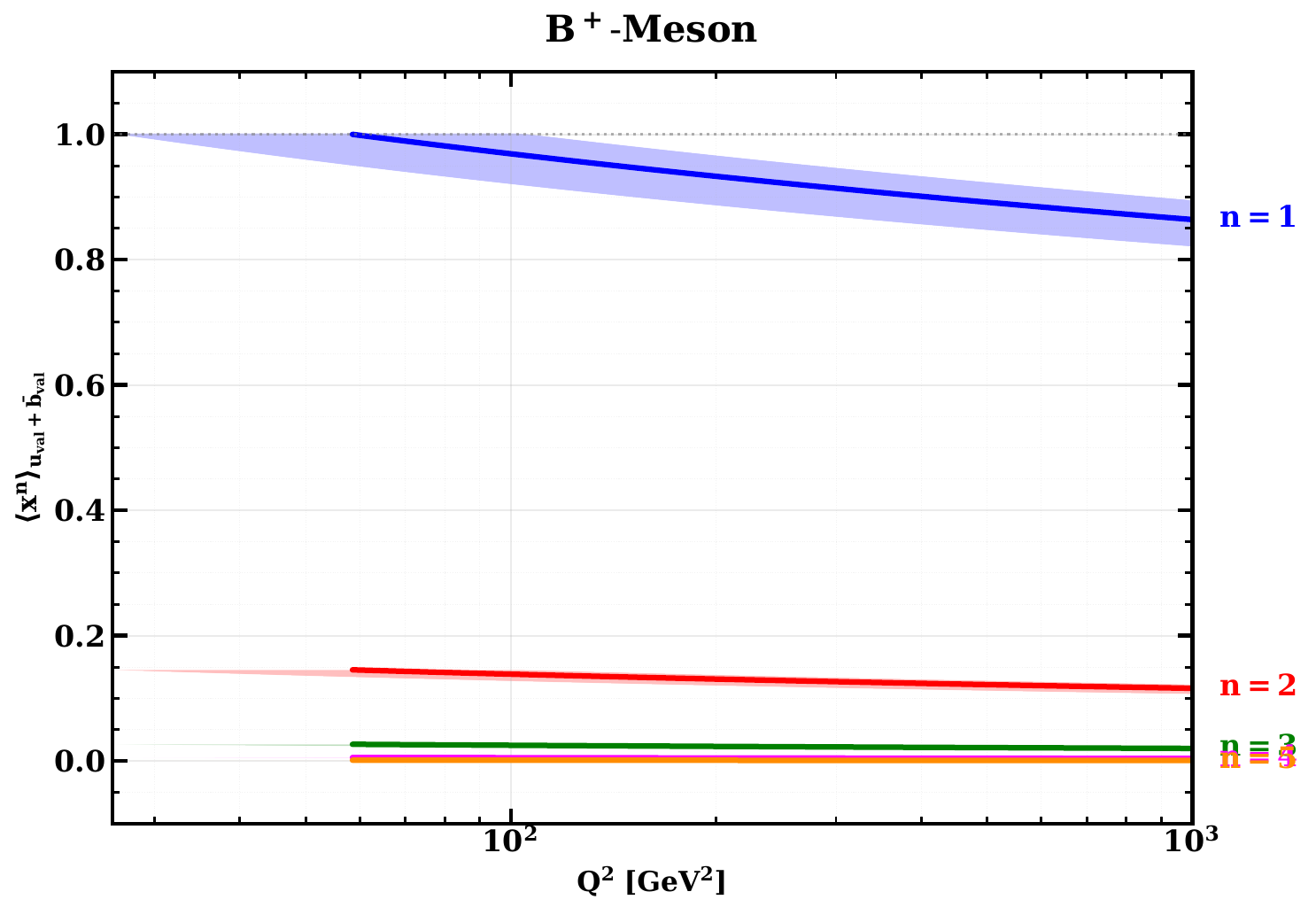}
\caption{The $n^{th}$ Mellin moment of the sum valence quark and antiquark $\langle x \rangle_{u_{val}+\bar b_{val}}$ with respect to $Q^2$ GeV$^2$ up to n=5.}
\end{subfigure}
\caption{(Color online) The $B^+(u \bar b)$-meson PDFs have been evolved to $Q^2=1000$ GeV$^2$ through NLO DGLAP evolutions by taking the initial scale $Q_0\in [5.10,10.20]$ GeV. The central line corresponds to the initial scale $Q_0= 7.65$ GeV.}
\label{bplus}
\end{figure*}

\bibliographystyle{apsrev}  
\bibliography{ref} 
\end{document}